\documentstyle[psfig]{mn_plus_bib}

\def\HST{{\it HST}}

\def\mnras{{MNRAS}}
\def\aj{{AJ}}

\def\apj{{ApJ}}
\def\apjs{{ApJS}}
\def\apjl{{ApJ Lett.}}
\def\aap{{A\&A}}
\def\araa{{ARA\&A}}

\def\pasp{{P.A.S.P.}}
\def\aap{{AAP}}
\def\aaps{{AAPS}}

\def\Msun{\hbox{M}_{\odot}}

\def\kms{\,\hbox{km}\,\hbox{s}^{-1}}

\def\H0{H_0=100 \, h \, {\rm kms^{-1}Mpc^{-1}}}

\def\etal{{et al.\thinspace}}
\def\eg{{e.g.\thinspace}}

\def\ie{{i.e.\thinspace}}

\def\V{{\emph{V}}}
\def\I{{\emph{I}}}

\def\R{{\emph{R}}}

\def\gsim{~\rlap{$>$}{\lower 1.0ex\hbox{$\sim$}}}

\begin{document}

\title{Keck Spectroscopy and Imaging of Globular Clusters in the Lenticular
Galaxy NGC~524}

\author[Beasley, Forbes, Brodie \& Kissler-Patig] {
  Michael A. Beasley$^{1}$\thanks{mbeasley@astro.swin.edu.au}, 
  Duncan A. Forbes$^{1}$, 
  Jean P. Brodie$^{2}$,
  Markus Kissler-Patig$^{3}$\\ \\
  $^1$ Astrophysics \& Supercomputing, Swinburne University,
  Hawthorn, VIC 3122, Australia\\
  $^2$ Lick Observatory, University of California,
 Santa Cruz, CA 95064, USA\\
  $^3$ European Southern Observatory, Karl-Schwarzschild-Str. 2,
85748 Garching, Germany\\
}

\date{Accepted~~~~~~~~~~.   Received~~~~~~~~~~.}

\pagerange{\pageref{firstpage}--\pageref{lastpage}}
\pubyear{2002}

\label{firstpage}

\maketitle

\begin{abstract}
We have obtained Keck LRIS imaging and spectra for 29 globular
clusters associated with the lenticular galaxy NGC~524.
Using the empirical calibration of Brodie \& Huchra we find 
that our spectroscopic sample spans a metallicity range of --2.0
$\leq$ [Fe/H] $\leq$ 0. 
We have compared the composite spectrum of the metal-poor 
([Fe/H] $<$ --1) and metal-rich clusters with stellar 
population models in order to estimate the ages of the NGC~524
globular clusters. 
We conclude that the clusters are generally old, and are
coeval at the 2$\sigma$ confidence level. 
To determine the mean [$\alpha$/Fe] ratios of the globular 
clusters, we have employed the Milone \etal $\alpha$-enhanced
stellar population models. We verified the reliability of these models 
by comparing them with high S/N Galactic globular cluster 
spectra. We observe a weak trend of decreasing [$\alpha$/Fe] ratios
with increasing metallicity in the NGC~524 clusters, 
the metal-poor cluster possess 
[$\alpha$/Fe]$\sim$0.3, whilst the metal-rich clusters exhibit
[$\alpha$/Fe] ratios closer to solar-scaled values.
Analysis of the cluster system kinematics reveals that the 
full sample (excluding an outlying cluster) exhibits a 
rotation of $114\pm60 \kms$ around
a position angle of $22\pm27 \deg$, and a velocity dispersion
of $186\pm29 \kms$ at a mean radius of 89$\arcsec$ 
from the galaxy centre.
Subdividing the clusters into metal-poor and metal-rich
subcomponents (at [Fe/H] = --1.0), we find that 
the metal-poor (17) clusters and metal-rich (11) clusters
have similar velocity dispersions ($197\pm40 \kms$
and $169\pm47 \kms$ respectively). However, the metal-poor
clusters dominate the rotation in our sample with 147$\pm75
\kms$, whilst the metal-rich clusters show no significant
rotation ($68\pm84 \kms$).
We derive a virial and projected mass estimation for NGC~524 of between
4 and 13$\times$ 10$^{11}$ $\Msun$ (depending on the assumed orbital
distribution) interior to $\sim$2 effective radii of this galaxy.
\end{abstract}


\section{Introduction}
\label{Introduction}


The spectroscopic study of globular cluster (GC) systems
has made great progress in the past few years. The availability
of 10 metre class telescopes, and efficient spectrographs, has meant
that it is now possible to obtain low-resolution spectra 
of individual GCs of sufficient quality to derive reliable
abundance and age information (in addition to valuable 
kinematic information) out to Virgo cluster distances.
However, such studies still only number a handful, and have
largely focused on the rich cluster systems of 
giant elliptical galaxies (\eg NGC~4486:\citeANP{Cohen98} 1998; 
NGC~1399:\citeANP{KisslerPatig98} 1998; 
 \citeANP{Forbes01} 2001; NGC~4472:\citeANP{Beasley00} 2000; 
\citeANP{Cohen03} 2003, see also \citeANP{Peng03} (2003) for a recent
study of NGC~5128).

Such studies have afforded unique insights into both the 
individual properties of GC systems associated with ellipticals,
and the relation between GCs and the formation of their host galaxies. 
The results of these previous works have shown that the 
GCs possess metallicities ranging from 1/400 to 
approximately solar values, with these most metal-rich 
GCs comparable to the integrated bulge starlight of 
their host galaxies. 
Moreover, within the uncertainties
(which remain necessarily model-dependent), 
many of GCs in these galaxies appear old and coeval.
Recent spectroscopic data (\eg \citeANP{Forbes01} 2001;
\citeANP{Larsen03} 2003) and IR imaging (\citeANP{KisslerPatig02}
2002; \citeANP{Puzia02} 2002)
have indicated the presence of young GCs in apparently otherwise 
undisturbed elliptical galaxies, suggesting complex
formation histories for at least a subset of ellipticals.

The situation for disk galaxies is even less well known.
Larsen \etal (2002) presented
a spectroscopic study of a small sample of
high S/N Sombrero galaxy (M104) GCs, and found that the clusters in this
relatively luminous (M$_B$ = --21.8) Sa spiral appear
old and coeval, similar to the Milky Way cluster
system. These authors found alpha-to-iron ratios
for the clusters of [$\alpha$/Fe] $\sim$ +0.4, typical of
luminous galaxy spheroids (\eg \citeANP{Trager98} 1998).
Recently, \citeANP{Kuntschner02} (2002) have obtained 
high-quality spectra for 17 GCs in the nearby ($\sim$ 10 Mpc) lenticular 
galaxy NGC~3115. 
Similar to the Sombrero GCs, the GCs in NGC~3115 generally
appear old, although interestingly,
\citeANP{Kuntschner02} (2002) were able to demonstrate a spread
in [$\alpha$/Fe] ratios amongst both the metal-rich and
metal-poor GCs, indicative of multiple phases of 
GC formation.

These latter results are particularly intriguing,
since we know that the Milky Way GCs are old
(i.e. $>$ 8 Gyr, \citeANP{Salaris02} 2002) 
and generally exhibit non-solar [$\alpha$/Fe] ratios
(\citeANP{Harris01} 2001).
Clearly few firm conclusions can be drawn from such
a small and restricted sample, and more information
is required with regards to late-type galaxies.

NGC~524 is an SA(rs)0 galaxy, dominating the small NGC~524 group
some 28.2 Mpc distant \cite{RC3}. Its GC system was first studied
in any detail by \citeANP{Harris85} (1985), who identified a
rich, and spatially extended system.
In fact it possesses one of the richest GC systems known for an 
S0 galaxy, with 4430 $\pm$ 950 GCs \cite{Harris91}.
More recently, the GC system of NGC~524 has been studied with \HST/WFPC2 by
\citeANP{Larsen01} (2001). From two pointings they identified a
total of 617 GCs to a magnitude limit of $\V \sim$ 26. Employing the KMM test
\cite{Ashman94} these authors determined the GC colour
distribution to be bimodal with peaks at $V-I$ = 0.98 and $V-I$ =
1.19 in the ratio $\sim$ 2:1. Using the colour-metallicity
relation of \citeANP{KisslerPatig98} (1998), these peaks correspond to
sub-populations with [Fe/H] $\sim$ --1.3 and --0.6.

In this paper, we present an analysis of the 
broad-band imaging and spectroscopy for
GC candidates associated with this galaxy,
the first such study to spectroscopically
investigate the GC system of NGC~524.

This paper is ordered in the following way:
in $\S$~\ref{Observations} we discuss
the data acquisition and reduction procedures.
Next, in $\S$~\ref{Indices}, we derive metallicities from
our integrated spectra, and compare line-strength
indices of the clusters to stellar population models.
In $\S$~\ref{Radial} we then investigate
the kinematical properties of the cluster
system and sub-populations, 
and derive dynamical  mass estimates for NGC~524.
Finally, we present a summary and our conclusions
in $\S$~\ref{Conclusions}.

\section{Observations and Data Reduction}
\label{Observations}

\subsection{Photometry}
\label{Photometry}

\begin{figure}
\centering
\centerline{\psfig{file=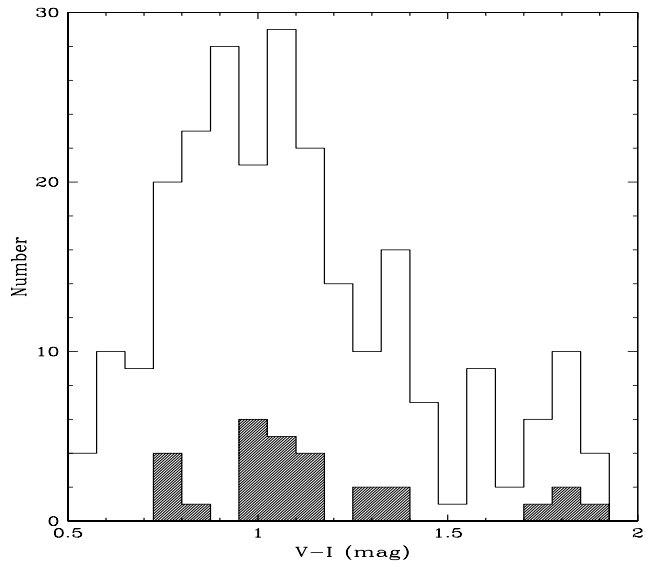,height=8cm,width=9cm}}
\caption{ The $V-I$ colour distribution of globular clusters
associated with NGC 524. The
open histogram shows the distribution for 245 candidate clusters
(after magnitude and colour selection) from Keck imaging. The
hashed histogram shows the colour distribution of {\it bona fide}
globular clusters for which we have obtained Keck spectra. These 
spectra sample the full range of colour of the NGC~524 globular 
cluster system. }
\label{fig:colours}
\end{figure}

Broad--band imaging of NGC~524 
in the $\V$, $\R$ and $\I$ filters was
obtained at the Keck-I telescope in 1996, September 8th
using the Low Resolution Imaging Spectrometer 
(LRIS; \citeANP{Oke95} 1995). The LRIS
instrument, equipped with a TEK $2048 \times 2048$ CCD, is 
mounted on the Cassegrain focus providing a
$0.215\,\mathrm{arcsec\,pixel^{-1}}$ imaging scale and a
$6^{\prime}\times7^{\prime}$ field--of--view. The total exposure 
times were 630 secs in $\V$, 330 secs in $\R$ and 300 secs in
$\I$. Seeing conditions were good, with a median
of $\sim$ 0.6 arcsec. These data were reduced following 
standard procedures, using IRAF
software. The reduced images were found to be flat to 
better than $\sim2\%$. Photometric calibration was performed 
using standard stars from \citeANP{Landolt92} (1992).

Selection of GC candidates from our LRIS multi-filter imaging 
was undertaken using the following selection criteria: 20.5 $< \V <$
24, 0.5 $< V-I <$ 2.0 and 0 $< V-R <$ 1.0. The magnitude
limits ensured that we selected the more luminous GCs 
(necessary for spectroscopy) whilst excluding bright foreground
stars. The colour cuts covered the full range expected for GCs including
photometric errors, whilst excluding extremely blue 
objects which were unlikely to be real GCs. 
After these selection criteria, we were left with a total
of 245 GC candidates.

In Figure~\ref{fig:colours} we show the colour distribution of 
our 245 GC candidates in NGC 524 (open histogram) and those for
which we have obtained LRIS spectra (filled histogram). 
The distribution shows a broad peak at $V-I \sim$ 1.0,
with possible enhancements at $V-I \sim$ 0.9 and 1.1,
similar to those seen in the HST data of \citeANP{Larsen01}
(2001). In an effort to obtain spectra of some very metal-rich
GCs, we also selected a number of candidates at $V-I \sim$ 1.8.
These turned out to be {\it bona fide} GCs, although not with the
high metallicities expected for these extremely red colours
(see Section~\ref{Metallicities}).

\begin{figure}
\centering
\centerline{\psfig{file=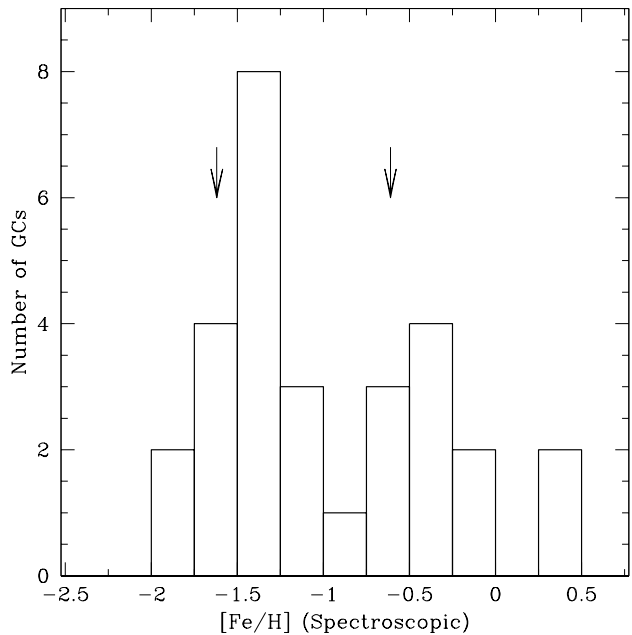,height=9cm}}
\caption{Metallicity distribution for our
spectroscopic sample of 29 NGC~524 GCs derived using the 
empirical calibration of Brodie \& Huchra
(1990). The two peaks of the distribution 
lie at approximately [Fe/H] $\sim$ --1.40 and [Fe/H] $\sim$ --0.40. 
The vertical arrows at [Fe/H] = --1.62 and --0.61 represent the
mean metallicities of the Galactic GC system after 
applying the KMM test to the Galactic GC catalogue of
Harris (1996) (see text).}
\label{fig:histogram}
\end{figure} 

\subsection{Spectroscopy}
\label{Spectroscopy}

\begin{figure}
\centering
\psfig{file=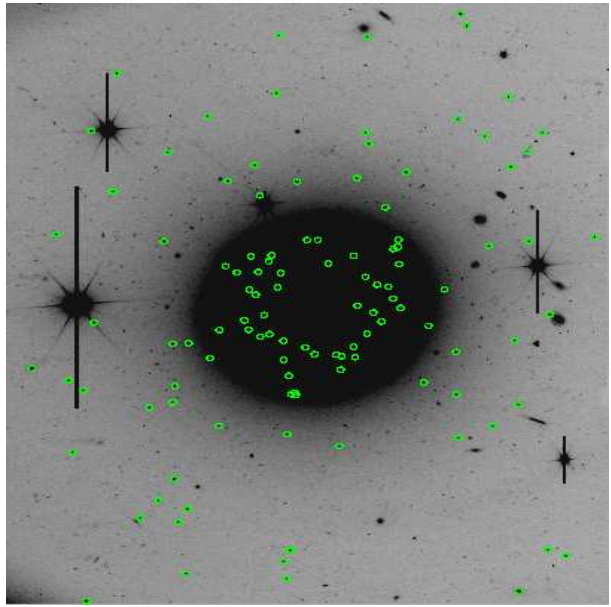,height=8cm,width=8cm}
\caption{Our globular cluster candidates (circles) for 
multi-object spectroscopy for both masks, shown on top of the
Keck LRIS V-band image (North up, East left). 
The LRIS field is approximately 6\arcmin$\times$7\arcmin .}
\label{fig:target}
\end{figure}

Spectra of GC candidates around NGC 524 were obtained with
LRIS on the Keck-I telescope. Candidate selection was based on the
imaging data, and was designed to cover a wide range of 
potential GC candidates.
Observations consisted of two masks
obtained on the nights of 1997 September 30th and October 1st.
The two masks used the same setup and had comparable exposure
times (\ie 10,800 and 9,000 sec). The 600~l/mm grating used
yielded a spectral resolution of 5.6\AA, with a full wavelength
range of 3640\AA\ -- 6200\AA. 

Data reduction was carried out using the {\tt REDUX} software
package developed by A. Phillips. Data reduction steps included
trimming these data, bias-subtraction, flat-fielding, removal of
any $x-$ and $y-$distortions, and producing optimally 
sky-subtracted, 1-D spectra. Comparison lamp spectra of Mercury,
Argon, Neon and Krypton were used for wavelength calibration. 
Spectra from the different nights were combined. Flux calibration was
provided by the flux standard BD~284211 observed on the first
night of the run. 

In Table~\ref{tab:candidates} we summarise the observational 
data for the objects for which we have obtained Keck spectra.
The $V$,$R$ \& $I$ photometry derived from our 1996 LRIS
observations were described in Section~\ref{Photometry}. 
Radial velocities and velocity uncertainties 
were measured from the spectra by
cross-correlating against high S/N M31 GC templates (225-280) and
(158-213) (using the nomenclature for the M31 GCs from
\citeANP{Huchra91} 1991 ), 
observed with the same instrumental set-up, using
\textsc{fxcor} in \textsc{iraf}. The zero-points of these
velocities were checked against 1.8\AA\ resolution synthetic 
spectra \cite{Vazdekis99}, broadened to the LRIS resolution. 
Of the 41 usable spectra obtained, 3 were Galactic stars, 9 were
background galaxies and 29 {\it bona fide} GCs. Thus our
magnitude and colour selection cuts have a contamination rate of
about 30\%.

\begin{table*}
\begin{center}
\renewcommand{\arraystretch}{1.2}
\caption{Globular candidates associated with NGC~524}
\begin{tabular}{lcccccll}
\hline
\hline
ID & R.A.   & Dec.    & $\V$    & $V-I$ & $V-R$ & $V_{\rm helio}$ & S/N\\
   &(J2000) & (J2000) & (mag) & (mag) & (mag) & (kms$^{-1}$) & (\AA$^{-1}$)\\
\hline
K004  &   1:24:34.71 &  9:30:14.0 & 21.77 & 1.03 & 0.41  & 2613$\pm$40&25\\
K005  &   1:24:35.29 &  9:31:09.0 & 21.90 & 1.26 & 0.58   & galaxy z=0.31&...\\
K010  &   1:24:37.16 &  9:30:09.0 & 21.46 & 1.52 & 0.73  & star&...\\
K014  &   1:24:39.22 &  9:30:06.3 & 21.21 & 1.73 & 0.84  & galaxy z=0.31&...\\
K015 &  1:24:40.50 &  9:29:08.5 &  22.43 & 1.03  & 0.41  & 2651$\pm$39&16\\
K018  &   1:24:41.37 &  9:31:10.1 & 21.93 & 1.03 & 0.39   & 2243$\pm$74&23\\
K023 &  1:24:42.89 &  9:30:04.9 & 22.39  & 0.75  & 0.44  & 2269$\pm$71&14\\
K024  &   1:24:43.28 &  9:31:15.0 & 21.88 & 1.13 & 0.50  & 2219$\pm$49&14\\
K029  &   1:24:43.68 &  9:30:06.1 & 22.16 & 1.16 & 0.47  & 2228$\pm$45&19\\
K031 &  1:24:43.95 &  9:29:05.8 & 21.83  & 0.60  & 0.23  & galaxy z=0.13&...\\
K033 &  1:24:44.49 &  9:31:40.5 & 21.92  & 1.80$\dag$  & 0.39  & 2775$\pm$90&10\\
K034 &  1:24:44.53 &  9:28:44.8 & 20.66  & 1.19  & 0.54  & galaxy z=0.34&...\\
K036 &  1:24:45.03 &  9:30:26.0 & 22.17  & 1.11  & 0.32  & 2380$\pm$53&15\\
K040 &  1:24:45.23 &  9:31:25.3 & 21.52  & 1.80$\dag$  & 0.39  & 3141$\pm$71&14\\
K043 &  1:24:45.59 &  9:31:47.8 &  21.92 & 1.05  & 0.44  & 2730$\pm$99&10\\
K044 &  1:24:45.71 &  9:30:04.6 & 22.01  & 0.75  & 0.44  & 2548$\pm$85&21\\
K051  &   1:24:46.38 &  9:30:31.0 & 22.40 & 0.27 & 0.32  & 2126$\pm$96&12\\
K052 &  1:24:46.40 &  9:30:47.8 & 21.31  & 0.76  & 0.19  & 2453$\pm$84&24\\
K054 &  1:24:46.73 &  9:29:19.9 & 21.84  & 1.85  & 0.68  & galaxy z=0.14&...\\
K056  &   1:24:46.84 &  9:30:45.3 & 21.72 & 0.77 & 0.54   & 2584$\pm$59&14\\
K057  &   1:24:47.11 &  9:30:56.6 & 20.93 & 1.39 & 0.51  & 2586$\pm$93&13\\
K058 &  1:24:47.17 &  9:33:39.4 & 21.02  & 1.84  & 0.80  & star&...\\
K061 &  1:24:47.57 &  9:31:49.8 & 20.77  & 1.73$\dag$  & 0.83  & 2629$\pm$63&12\\
K065 &  1:24:48.36 &  9:32:39.4 & 21.67  & 1.16  & 0.45  & 2500$\pm$23&25\\
K067 &  1:24:48.49 &  9:32:07.3 & 21.73  & 1.28  & 0.44  & 2230$\pm$82&11\\
K069 &  1:24:48.95 &  9:31:54.3 & 20.97  & 1.33  & 0.56  & 2465$\pm$110&10\\
K074 &  1:24:49.46 &  9:30:34.6 & 21.72  & 1.02  & 0.40  & 2702$\pm$66&21\\
K075 &  1:24:49.58 &  9:32:13.5 & 21.49  & 0.94  & 0.49  & galaxy z=0.06&...\\
K076  &   1:24:49.59 &  9:31:33.2 & 20.77 & 1.43 & 0.39  & galaxy z=0.06&...\\
K078  &   1:24:49.93 &  9:30:49.1 & 21.54 & 0.99 & 0.33   & 2644$\pm$106&12\\
K083 &  1:24:50.45 &  9:33:04.8 & 22.22  & 1.00  & 0.45  & 2498$\pm$52&16\\
K085 &  1:24:50.70 &  9:33:27.4 & 22.15  & 1.02  & 0.44  & 2453$\pm$58&17\\
K086  &   1:24:50.72 &  9:31:20.9 & 21.59 & 1.32 & 0.48  & 2375$\pm$41&20\\
K089 &  1:24:50.90 &  9:34:05.1 & 21.25  & 1.02  & 0.47  & 2577$\pm$36&23\\
K092  &   1:24:52.87 &  9:30:54.1 & 21.56 & 1.88$\dag$ & 0.95  & 2131$\pm$130&11\\
K099  &   1:24:54.25 &  9:33:17.2 & 21.75 & 0.68 & 0.30   & galaxy z=0.19&...\\
K104 &  1:24:55.72 &  9:33:02.6 & 22.24  & 1.07  & 0.43  & galaxy z=0.06&...\\
K109  &   1:24:56.68 &  9:30:24.6 & 22.13 & 1.03 & 0.39   & 2183$\pm$86&14\\
K110  &   1:24:57.58 &  9:33:15.8 & 22.18 & 0.85 & 0.38  & 2115$\pm$107&19\\
K113  &   1:25:00.47 &  9:31:55.7 & 22.33 & 0.99 & 0.45  & 2091$\pm$83&11\\
K115  &   1:25:00.99 &  9:32:51.9 & 21.46 & 0.83 & 0.28  & star&...\\
\hline
\end{tabular}
\label{tab:candidates}
\end{center}
$\dag$ : suspect photometry due to dust.
\end{table*} 

Line-strength indices both in the Lick system 
(\eg \citeANP{Trager98} 1998 and references therein)
and using the \citeANP{Brodie90} (1990) definitions
were measured from our flux-calibrated spectra. 
Due to the variable nature
of the wavelength ranges in multi-slit spectra, the same set
of indices were not measured for all spectra.
Uncertainties in the indices were derived from the 
photon noise in the un-fluxed spectra following
\citeANP{Cardiel98} (1998). In general these were in good agreement
with the uncertainties obtained from a Monte Carlo approach
(\eg \citeANP{Kuntschner02} 2002). However, a few of the lowest
S/N spectra yielded Monte Carlo errors which where significantly 
larger than the \citeANP{Cardiel98} (1998) prescription,
suggesting that in this regime other effects (such as
flat-fielding) dominate the index uncertainties.

Our resulting calibrated spectra possessed S/N = 10--25 \AA$^{-1}$, 
corresponding to errors in the \eg H$\beta$ index of 0.9--0.3
\AA. 

\section{Spectroscopic Indices of the GCs}
\label{Indices}

\subsection{Metallicities}
\label{Metallicities}

There were two approaches we considered in estimating
the metallicities of the NGC~524 GCs from our integrated
spectra. The empirical calibration of \citeANP{Brodie90} (1990)
was specifically designed to estimate metallicities for
extragalactic GCs from low-resolution, and potentially 
low S/N spectra.
Alternatively, stellar population models may be employed
to derive metallicities either by assuming an age for
the GCs (necessary due to the age--metallicity degeneracy), or
by allowing age discrimination to come from Balmer indices.

The former technique is tied to Milky Way and Andromeda GC 
calibrators with independently derived metallicities, and has
been employed with some success by a number of workers (\eg
\citeANP{Brodie91} 1991; \citeANP{Perelmuter95} 1995;
\citeANP{Bridges97} 1997; \citeANP{KisslerPatig98} 1998;
\citeANP{Barmby00} 2000; \citeANP{Schroder02} 2002).
This is, however, also an important drawback;
the scarcity of solar, and lack of super-solar calibrators makes 
the metal-rich end poorly constrained (\eg \citeANP{KisslerPatig98} 1998).
The latter method has the great benefit of employing 
input stellar libraries based on both halo and disk
stars which in principle extend to above solar values.
However, the key disadvantage in these models is that, 
whilst halo stars are known to exhibit $\alpha$-element
enhancement similar to Galactic GCs of all metallicities
(\eg \citeANP{Harris01} 2001), disk stars
have scaled-solar abundances. 
This has complicated the interpretation
of stellar population models for metallicity (and age)
derivations. Moreover, the use of individual indices coupled with
the Balmer lines generally requires reasonably high S/N data.
In view of these issues, we have initially adopted the
\citeANP{Brodie90} (1990) calibration to derive metallicities for
the NGC~524 GCs. Later, we examine the ages and abundance ratios of the NGC~524
GCs using SSP models.

\begin{table*}
\begin{center}
\renewcommand{\arraystretch}{1.2}
\caption{Raw index measurements of metallicity indices as
prescribed in Brodie \& Huchra (1990). }
\begin{tabular}{lccccccc}
\hline
\hline
ID & CNR & G band & MgG & MgH & Mg2 & MgB & Fe52 \\
 & (\AA) & (\AA) & (\AA) & (\AA) & (mag) & (\AA) & (\AA)\\ 
\hline
K004	&	1.09$\pm$0.57	&	3.13$\pm$0.56	&	1.03$\pm$0.36	&	1.05$\pm$0.39	&	0.063$\pm$0.008	&	1.38$\pm$0.30	&	1.28$\pm$0.34\\
K015	&	0.51$\pm$0.76	&	5.18$\pm$0.75	&	1.19$\pm$0.52	&	0.80$\pm$0.55	&	0.064$\pm$0.011	&	1.80$\pm$0.43	&	1.26$\pm$0.49\\
K018	&	--4.25$\pm$0.67	&	3.22$\pm$0.58	&	--1.31$\pm$0.41	&	1.42$\pm$0.42	&	--0.008$\pm$0.008	&	--0.19$\pm$0.34	&	0.24$\pm$0.38\\
K023	&	1.63$\pm$0.73	&	2.93$\pm$0.77	&	1.90$\pm$0.54	&	0.30$\pm$0.59	&	0.067$\pm$0.012	&	1.66$\pm$0.46	&	1.66$\pm$0.51\\
K024	&	5.42$\pm$0.68	&	6.01$\pm$0.76	&	0.42$\pm$0.54	&	7.88$\pm$0.51	&	0.206$\pm$0.011	&	2.71$\pm$0.48	&	3.59$\pm$0.47\\
K029	&	3.48$\pm$0.67	&	7.00$\pm$0.63	&	--0.25$\pm$0.45	&	4.10$\pm$0.44	&	0.298$\pm$0.009	&	5.24$\pm$0.38	&	2.95$\pm$0.40\\
K033	&	0.64$\pm$0.85	&	1.98$\pm$0.91	&	3.30$\pm$0.75	&	--0.65$\pm$0.84	&	0.136$\pm$0.017	&	--0.17$\pm$0.70	&	2.26$\pm$0.76\\
K036	&	9.97$\pm$0.58	&	5.68$\pm$0.76	&	0.61$\pm$0.56	&	6.30$\pm$0.53	&	0.306$\pm$0.012	&	4.78$\pm$0.49	&	4.82$\pm$0.49\\
K040	&	6.15$\pm$0.60	&	5.19$\pm$0.68	&	2.68$\pm$0.52	&	1.50$\pm$0.56	&	0.163$\pm$0.011	&	1.56$\pm$0.44	&	3.29$\pm$0.50\\
K043	&	--2.07$\pm$1.00	&	6.73$\pm$1.12	&	0.37$\pm$1.03	&	3.93$\pm$0.99	&	0.257$\pm$0.021	&	10.32$\pm$0.57	&	7.44$\pm$0.85\\
K044	&	--2.71$\pm$0.60	&	2.79$\pm$0.56	&	0.96$\pm$0.42	&	0.67$\pm$0.45	&	0.049$\pm$0.009	&	0.69$\pm$0.35	&	1.39$\pm$0.40\\
K051	&	0.83$\pm$0.79	&	2.89$\pm$0.83	&	0.37$\pm$0.66	&	1.02$\pm$0.72	&	0.084$\pm$0.014	&	2.76$\pm$0.58	&	2.64$\pm$0.65\\
K052	&	--2.04$\pm$0.52	&	2.97$\pm$0.49	&	1.62$\pm$0.38	&	0.51$\pm$0.42	&	0.027$\pm$0.008	&	1.79$\pm$0.31	&	--0.39$\pm$0.39\\
K056	&	--0.15$\pm$0.73	&	3.12$\pm$0.66	&	--3.73$\pm$0.63	&	3.34$\pm$0.58	&	0.054$\pm$0.012	&	--2.24$\pm$0.52	&	--0.22$\pm$0.57\\
K057	&	7.35$\pm$0.90	&	3.69$\pm$0.76	&	3.72$\pm$0.57	&	2.17$\pm$0.61	&	0.233$\pm$0.013	&	4.99$\pm$0.48	&	2.28$\pm$0.57\\
K061	&	--1.19$\pm$0.67	&	0.81$\pm$1.05	&	1.38$\pm$0.66	&	1.22$\pm$0.78	&	0.090$\pm$0.015	&	3.19$\pm$0.52	&	0.67$\pm$0.70\\
K065	&	1.94$\pm$0.49	&	6.49$\pm$0.47	&	1.03$\pm$0.38	&	2.47$\pm$0.39	&	0.216$\pm$0.008	&	3.54$\pm$0.34	&	3.93$\pm$0.34\\
K067	&	--6.22$\pm$1.00	&	1.90$\pm$1.09	&	3.51$\pm$0.79	&	4.31$\pm$0.86	&	0.074$\pm$0.018	&	1.58$\pm$0.71	&	5.81$\pm$0.75\\
K069	&	--5.61$\pm$0.97	&	7.68$\pm$0.91	&	6.54$\pm$0.69	&	0.61$\pm$0.78	&	0.200$\pm$0.017	&	3.67$\pm$0.58	&	2.61$\pm$0.71\\
K074	&	--1.54$\pm$0.59	&	3.58$\pm$0.55	&	2.04$\pm$0.42	&	1.62$\pm$0.45	&	0.118$\pm$0.009	&	1.65$\pm$0.35	&	0.20$\pm$0.42\\
K078	&	--1.90$\pm$0.99	&	5.51$\pm$1.00	&	5.64$\pm$0.76	&	0.14$\pm$0.91	&	0.119$\pm$0.018	&	--1.62$\pm$0.72	&	7.81$\pm$0.79\\
K083	&	4.10$\pm$0.71	&	6.35$\pm$0.71	&	0.57$\pm$0.54	&	1.85$\pm$0.55	&	0.174$\pm$0.011	&	3.63$\pm$0.45	&	1.54$\pm$0.52\\
K085	&	4.34$\pm$0.63	&	1.76$\pm$0.73	&	--0.12$\pm$0.50	&	0.40$\pm$0.52	&	0.129$\pm$0.010	&	1.65$\pm$0.43	&	1.93$\pm$0.46\\
K086	&	0.05$\pm$0.59	&	4.88$\pm$0.56	&	2.72$\pm$0.41	&	0.59$\pm$0.46	&	0.092$\pm$0.009	&	1.93$\pm$0.35	&	0.45$\pm$0.41\\
K089	&	1.09$\pm$0.54	&	5.34$\pm$0.52	&	2.52$\pm$0.38	&	1.16$\pm$0.41	&	0.140$\pm$0.008	&	2.26$\pm$0.32	&	2.03$\pm$0.35\\
K092	&	--8.44$\pm$1.07	&	1.26$\pm$0.96	&	4.61$\pm$0.64	&	0.29$\pm$0.75	&	0.116$\pm$0.015	&	0.58$\pm$0.61	&	0.90$\pm$0.74\\
K109	&	9.73$\pm$0.49	&	3.29$\pm$0.67	&	--2.70$\pm$0.62	&	3.75$\pm$0.56	&	0.083$\pm$0.012	&	0.31$\pm$0.53	&	3.45$\pm$0.56\\
K110	&	--1.95$\pm$0.78	&	4.70$\pm$0.71	&	1.58$\pm$0.44	&	--0.89$\pm$0.50	&	0.050$\pm$0.010	&	0.22$\pm$0.39	&	0.48$\pm$0.42\\
K113	&	--3.69$\pm$1.06	&	5.48$\pm$0.90	&	7.23$\pm$0.63	&	3.37$\pm$0.72	&	0.259$\pm$0.015	&	2.02$\pm$0.59	&	3.31$\pm$0.63\\

\hline
\end{tabular}
\label{tab:RawBrodie}
\end{center}
\end{table*}

To employ the metallicity calibration of \citeANP{Brodie90} (1990), 
we have measured the line--strength indices of the NGC~524 GCs
using the definitions given in their paper. 
For our metallicity determinations we have used 
seven indices calibrated by \citeANP{Brodie90} (1990), these
are CNR, G band, MgG, MgH, Mg2, MgB and Fe52. 
We did not use indices bluewards of 
CNR (\eg CNB, $\Delta$, H+K), since the majority of the
spectra do not reach to these shorter wavelengths. 
Moreover, we did not include Na D ($\lambda\lambda$ 5890 \AA)
since this index was severely affected by sodium sky lines 
(see \citeANP{Larsen02a} 2002).
We have opted to leave our data in their 'natural' resolution
($\sim$ 5.6\AA). Any systematic differences due to the
differing resolutions of these data and those of
\citeANP{Brodie90} (1990) are expected to be small 
(e.g. \citeANP{Larsen02a} 2002).

These 'raw' index measurements of the NGC~524 GCs are given in 
Table~\ref{tab:RawBrodie} along with their associated photon error.
The individual and mean metallicities for the NGC~524 GCs using
the \citeANP{Brodie90} (1990) calibration are listed 
in Table~\ref{tab:metallicities}.
The uncertainties on these mean metallicities were determined using 
the correction to the mean metallicity estimates as discussed 
in \citeANP{Larsen02a} (2002). 
Note that since the \citeANP{Brodie90} (1990) calibration 
is tied to the \citeANP{Zinn84} (1984) scale, all metallicities
derived using this calibration are given in \citeANP{Zinn84}'s
(1984) [Fe/H]. 

\begin{table*}
\begin{center}
\renewcommand{\arraystretch}{1.2}
\caption{Individual and final weighted mean metallicities (on the Zinn \& West (1984) scale) for the 
NGC~524 GCs using the Brodie \& Huchra (1990) calibration.}
\begin{tabular}{lcccccccc}
\hline
\hline
ID & CNR & G band & MgG & MgH & Mg2 & MgB & Fe52 & $\langle$[Fe/H]$\rangle$\\
\hline
K004	&	--0.92$\pm$0.47	&	--1.25$\pm$0.38	&	--1.67$\pm$0.43	&	--1.47$\pm$0.50	&	--1.58$\pm$0.35	&	--1.52$\pm$0.51	&	--1.34$\pm$0.64	&	--1.40$\pm$0.08\\
K015	&	--1.07$\pm$0.49	&	--0.39$\pm$0.42	&	--1.59$\pm$0.46	&	--1.56$\pm$0.52	&	--1.57$\pm$0.35	&	--1.30$\pm$0.53	&	--1.35$\pm$0.67	&	--1.25$\pm$0.15\\
K018	&	--2.16$\pm$0.48	&	--1.21$\pm$0.38	&	--2.70$\pm$0.44	&	--1.34$\pm$0.50	&	--2.29$\pm$0.35	&	--2.31$\pm$0.52	&	--1.95$\pm$0.64	&	--1.99$\pm$0.18\\
K023	&	--0.79$\pm$0.48	&	--1.33$\pm$0.43	&	--1.27$\pm$0.47	&	--1.74$\pm$0.52	&	--1.54$\pm$0.36	&	--1.38$\pm$0.54	&	--1.12$\pm$0.67	&	--1.32$\pm$0.10\\
K024	&	0.24$\pm$0.48	&	--0.03$\pm$0.42	&	--1.94$\pm$0.47	&	1.11$\pm$0.51	&	--0.16$\pm$0.35	&	--0.82$\pm$0.54	&	0.07$\pm$0.66	&	--0.23$\pm$0.31\\
K029	&	--0.30$\pm$0.48	&	0.43$\pm$0.39	&	--2.24$\pm$0.45	&	--0.35$\pm$0.50	&	0.75$\pm$0.35	&	0.61$\pm$0.53	&	--0.34$\pm$0.65	&	--0.16$\pm$0.34\\
K033	&	--1.03$\pm$0.50	&	--1.71$\pm$0.46	&	--0.61$\pm$0.53	&	--2.07$\pm$0.56	&	--0.86$\pm$0.38	&	--2.30$\pm$0.60	&	--0.76$\pm$0.75	&	--1.31$\pm$0.24\\
K036	&	1.68$\pm$0.47	&	--0.17$\pm$0.42	&	--1.85$\pm$0.48	&	0.49$\pm$0.52	&	0.83$\pm$0.36	&	0.34$\pm$0.55	&	0.85$\pm$0.67	&	0.30$\pm$0.37\\
K040	&	0.45$\pm$0.47	&	--0.39$\pm$0.40	&	--0.90$\pm$0.47	&	--1.31$\pm$0.52	&	--0.59$\pm$0.35	&	--1.43$\pm$0.54	&	--0.12$\pm$0.67	&	--0.61$\pm$0.21\\
K043	&	--1.68$\pm$0.51	&	0.30$\pm$0.52	&	--1.97$\pm$0.61	&	--0.42$\pm$0.59	&	0.34$\pm$0.40	&	3.95$\pm$0.57	&	2.63$\pm$0.78	&	0.37$\pm$0.80\\
K044	&	--1.82$\pm$0.47	&	--1.39$\pm$0.38	&	--1.70$\pm$0.44	&	--1.60$\pm$0.50	&	--1.72$\pm$0.35	&	--1.87$\pm$0.52	&	--1.28$\pm$0.65	&	--1.63$\pm$0.07\\
K051	&	--0.99$\pm$0.49	&	--1.35$\pm$0.44	&	--1.97$\pm$0.50	&	--1.48$\pm$0.54	&	--1.38$\pm$0.36	&	--0.79$\pm$0.57	&	--0.52$\pm$0.71	&	--1.25$\pm$0.16\\
K052	&	--1.67$\pm$0.47	&	--1.31$\pm$0.36	&	--1.39$\pm$0.44	&	--1.66$\pm$0.50	&	--1.94$\pm$0.35	&	--1.31$\pm$0.52	&	--2.31$\pm$0.64	&	--1.63$\pm$0.11\\
K056	&	--1.23$\pm$0.48	&	--1.25$\pm$0.40	&	--3.72$\pm$0.49	&	--0.64$\pm$0.52	&	--1.67$\pm$0.36	&	--3.28$\pm$0.55	&	--2.21$\pm$0.69	&	--1.94$\pm$0.37\\
K057	&	0.82$\pm$0.50	&	--1.02$\pm$0.42	&	--0.40$\pm$0.48	&	--1.07$\pm$0.53	&	0.10$\pm$0.36	&	0.46$\pm$0.55	&	--0.74$\pm$0.69	&	--0.25$\pm$0.25\\
K061	&	--1.48$\pm$0.48	&	--2.15$\pm$0.50	&	--1.51$\pm$0.50	&	--1.41$\pm$0.55	&	--1.32$\pm$0.37	&	--0.56$\pm$0.55	&	--1.70$\pm$0.73	&	--1.44$\pm$0.17\\
K065	&	--0.71$\pm$0.47	&	0.19$\pm$0.36	&	--1.66$\pm$0.44	&	--0.96$\pm$0.50	&	--0.07$\pm$0.35	&	--0.37$\pm$0.52	&	0.28$\pm$0.63	&	--0.46$\pm$0.22\\
K067	&	--2.57$\pm$0.51	&	--1.73$\pm$0.51	&	--0.51$\pm$0.54	&	--0.27$\pm$0.57	&	--1.47$\pm$0.38	&	--1.42$\pm$0.60	&	1.50$\pm$0.74	&	--1.07$\pm$0.45\\
K069	&	--2.44$\pm$0.51	&	0.75$\pm$0.46	&	1.01$\pm$0.51	&	--1.62$\pm$0.55	&	--0.22$\pm$0.38	&	--0.29$\pm$0.57	&	--0.55$\pm$0.73	&	--0.44$\pm$0.44\\
K074	&	--1.56$\pm$0.47	&	--1.07$\pm$0.37	&	--1.20$\pm$0.45	&	--1.26$\pm$0.50	&	--1.04$\pm$0.35	&	--1.38$\pm$0.52	&	--1.97$\pm$0.65	&	--1.30$\pm$0.10\\
K078	&	--1.64$\pm$0.51	&	--0.25$\pm$0.49	&	0.55$\pm$0.53	&	--1.79$\pm$0.58	&	--1.03$\pm$0.38	&	--3.00$\pm$0.61	&	2.89$\pm$0.76	&	--0.73$\pm$0.63\\
K083	&	--0.13$\pm$0.48	&	0.13$\pm$0.41	&	--1.87$\pm$0.47	&	--1.18$\pm$0.52	&	--0.48$\pm$0.35	&	--0.31$\pm$0.54	&	--1.19$\pm$0.67	&	--0.67$\pm$0.24\\
K085	&	--0.07$\pm$0.48	&	--1.79$\pm$0.41	&	--2.18$\pm$0.46	&	--1.70$\pm$0.51	&	--0.93$\pm$0.35	&	--1.38$\pm$0.54	&	--0.96$\pm$0.66	&	--1.29$\pm$0.23\\
K086	&	--1.18$\pm$0.47	&	--0.52$\pm$0.38	&	--0.88$\pm$0.44	&	--1.63$\pm$0.51	&	--1.30$\pm$0.35	&	--1.24$\pm$0.52	&	--1.83$\pm$0.65	&	--1.17$\pm$0.14\\
K089	&	--0.92$\pm$0.47	&	--0.32$\pm$0.37	&	--0.98$\pm$0.44	&	--1.43$\pm$0.50	&	--0.82$\pm$0.35	&	--1.06$\pm$0.52	&	--0.90$\pm$0.64	&	--0.89$\pm$0.11\\
K092	&	--3.01$\pm$0.52	&	--1.98$\pm$0.47	&	0.03$\pm$0.50	&	--1.74$\pm$0.55	&	--1.06$\pm$0.37	&	--1.93$\pm$0.58	&	--1.56$\pm$0.74	&	--1.57$\pm$0.34\\
K109	&	1.59$\pm$0.47	&	--1.18$\pm$0.40	&	--3.29$\pm$0.49	&	--0.49$\pm$0.52	&	--1.38$\pm$0.36	&	--2.06$\pm$0.56	&	--0.03$\pm$0.69	&	--1.01$\pm$0.51\\
K110	&	--1.65$\pm$0.49	&	--0.60$\pm$0.41	&	--1.41$\pm$0.45	&	--2.15$\pm$0.51	&	--1.71$\pm$0.35	&	--2.11$\pm$0.53	&	--1.81$\pm$0.65	&	--1.60$\pm$0.18\\
K113	&	--2.04$\pm$0.52	&	--0.26$\pm$0.46	&	1.37$\pm$0.49	&	--0.63$\pm$0.54	&	0.36$\pm$0.37	&	--1.19$\pm$0.57	&	--0.11$\pm$0.70	&	--0.30$\pm$0.39\\
\hline
\end{tabular}
\label{tab:metallicities}
\end{center}
\end{table*}

\begin{figure}
\centering
\centerline{\psfig{file=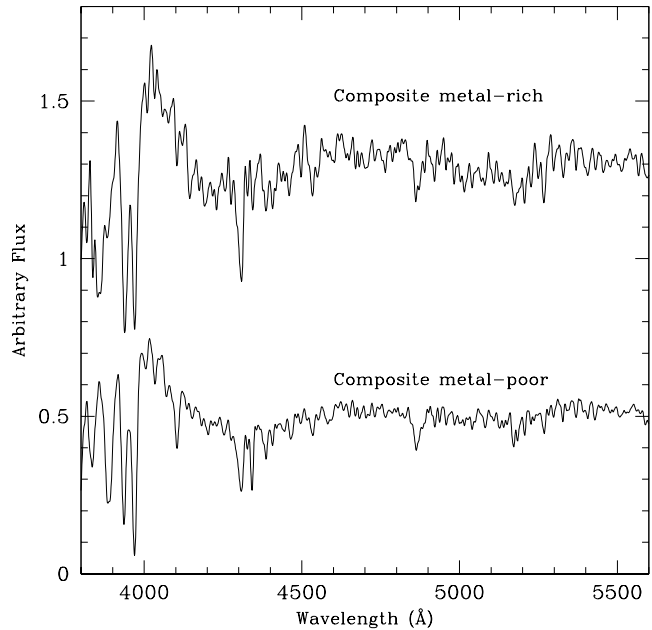,height=9cm}}
\caption{Combined spectra of metal-rich ([Fe/H]$\geq$--1.0) 
and metal-poor ([Fe/H]$<$--1.0) NGC~524 GCs. The spectra
have been smoothed to the Lick resolution (8--11 \AA\ FWHM) and
normalised using a low-order polynomial. A constant has been 
added to the metal-rich spectrum for display purposes.}
\label{fig:specs}
\end{figure} 

We show a histogram of the mean spectroscopic metallicities
derived for the NGC~524 GCs (last column of 
Table~\ref{tab:metallicities}) in Figure~\ref{fig:histogram}. 
Using the \citeANP{Brodie90} (1990) calibration, 
our sample of 29 GCs spans a metallicity range of --2.0 $\leq$
[Fe/H] $\leq$ 0.4. 
The high metallicities of the two most metal-rich GCs
shown in Figure~\ref{fig:histogram} are possibly
overestimates. These are our lowest S/N spectra (reflected
in their large uncertainties in Table~\ref{tab:metallicities}),
and the \citeANP{Brodie90} (1990) calibration for Mg2 and
MgH over-predicts [Fe/H] (with respect to stellar population models)
at high metallicities \cite{KisslerPatig98}.
The distribution appears bimodal
with peaks at approximately [Fe/H] $\sim$ --1.40 and [Fe/H] 
$\sim$ --0.4. However, not only is the significance of this 
bimodality rather low due to small number statistics, but also
there is a bias inherent in our spectroscopic sample since we deliberately
targeted metal-poor and metal-rich clusters for study.
An unweighted mean of the metallicities shown in
Figure~\ref{fig:histogram} yields a mean metallicity 
of [Fe/H]=--0.97, with a dispersion of 0.62 dex.
 
The two vertical arrows indicate the positions of the 
mean metallicities of the metal-poor and metal-rich Milky Way 
GC subpopulations at [Fe/H]=--1.62 and --0.61. These were
derived by applying the homoscedastic KMM test \cite{Ashman94}
to the GC metallicities in the February 2003 version of the 
\citeANP{Harris96} (1996) catalogue\footnote{http://physun.physics.mcmaster.ca/~harris/mwgc.dat}.
From Figure~\ref{fig:histogram} alone we may conclude the
metallicity distribution of the NGC~524 GCs is similar to
that of the Milky Way, although the metallicity distribution
for the NGC~524 GCs appears $\sim$ 0.15 dex more metal-rich.
We also appear to be missing some of the most metal-poor GCs seen in
the Galaxy, which is probably a result of our colour selection criteria.

Four of our spectral sample of GCs (K033, K040, K061 \& 
K092) possess very red colours (V--I $>$ 1.7), and were
selected from our initial catalogue as being candidate 
high-metallicity GCs. 
However, whilst our radial velocity measurements confirm 
their GC status, the spectroscopic metallicities of these 
GCs indicate that they have significantly less than 
solar metallicity (e.g. see Table~\ref{tab:metallicities}).
Their spectroscopic metallicities are in disagreement  
with empirical colour-metallicity relations, the 
flattest of which \cite{KisslerPatig98} predicts 
[Fe/H] $\geq$ +1.0 for such red colours.

All of these GCs lie within 1 arcminute of the galaxy centre,  
and a plot of $V-I$ colour versus galactocentric radius for the
full Keck sample reveals a mean $V-I$ of $\sim$1.0, 
except interior to 1 arcminute, when the
mean colour jumps to $V-I \sim$1.5.
This suggests that either reddening by dust, 
or possibly the high galaxy background at small radii is 
affecting the $V-I$ colours.

\citeANP{Silchenko00} (2000) identified a slightly inclined 
(by $\sim$ 20 degrees) ring of dust and gas
in the centre of NGC~524, approximately 20 arcseconds ($\sim$
3 kpc) in radius. This encompasses the projected radius of 
the innermost GC (K061), with K033 and K040 in close proximity.
GC K092 is slightly more distant, at a projected 
radius of $\sim$ 50 arcseconds ($\sim$ 8 kpc) from the 
centre of NGC~524. We thus conclude that the very red $V-I$ colours of
these four GCs are likely affected by dust. This finding emphasises the 
importance of spectroscopic indices in determining accurate 
metallicities for GCs in galaxies likely to host
significant amounts of dust.

\subsection{Metallicity and Age Constraints using Stellar Population Models}
\label{Ages}

It can be seen in Table~\ref{tab:metallicities} that the
uncertainties in the individual indices for the majority of
the GCs are relatively large, which impacts severely on our ability to
constrain ages and abundance ratios for the GCs. Therefore, in addition
to investigating individual clusters, we
have co-added the NGC~524 GCs into a metal-rich and metal-poor
``composite'' GC, taking the separation between sub-populations as
[Fe/H] = --1, as indicated by \eg Figure~\ref{fig:histogram}.
The spectra of our composite metal-rich and metal-poor GCs
are shown in Figure~\ref{fig:specs}. 

In order to examine the metallicities and ages of the NGC~524 GCs,
we have compared Lick indices measured for the NGC~524 GCs
with simple stellar population (SSP) models.
We measured Lick indices for the individual and 
composite NGC~524 GCs using the index definitions given in
\citeANP{Trager98} (1998). These have been  
supplemented by measurements of the higher-order 
H$\gamma$ and H$\delta$ Balmer lines as 
defined by \citeANP{WO97} (1997).
To provide a first-order correction to the Lick system
resolution, we have broadened our data to a FWHM of 
8--11 \AA\ using a wavelength-dependent Gaussian kernel.
The Lick indices and the respective uncertainties 
for the NGC~524 GCs are listed in Table~\ref{tab:Lick}.

\begin{table*}
\begin{center}
\renewcommand{\arraystretch}{1.2}
\caption{Lick indices of the NGC~524 GCs.}
\begin{tabular}{lccccccc}
\hline
\hline
ID & Mg$_2$ & Mg $b$ & Fe5270 & Fe5335 & H$\beta$ & H$\gamma_A$ &
H$\delta_A$\\
 & (mag) & (\AA) & (\AA) & (\AA) & (\AA) & (\AA) & (\AA)\\ 
\hline
K004 & 0.066 $\pm$ 0.008 & 1.70 $\pm$ 0.30 & 1.36 $\pm$ 0.34 & 1.86 $\pm$ 0.39 & 2.42 $\pm$ 0.30 & --0.90 $\pm$ 0.56 & 5.38 $\pm$ 0.63\\
K015 & 0.067 $\pm$ 0.011 & 2.22 $\pm$ 0.43 & 1.31 $\pm$ 0.49 & 1.80 $\pm$ 0.55 & 2.47 $\pm$ 0.43 & --3.29 $\pm$ 0.83 & 0.99 $\pm$ 0.86\\
K018 & --0.005 $\pm$ 0.008 & 0.13 $\pm$ 0.35 & 0.29 $\pm$ 0.39 & 1.60 $\pm$ 0.43 & 2.83 $\pm$ 0.32 & --3.69 $\pm$ 0.62 & 5.48 $\pm$ 0.63\\
K023 & 0.070 $\pm$ 0.012 & 1.88 $\pm$ 0.46 & 1.76 $\pm$ 0.52 & 4.04 $\pm$ 0.56 & 3.81 $\pm$ 0.41 & --2.39 $\pm$ 0.82 & 1.41 $\pm$ 0.84\\
K024 & 0.209 $\pm$ 0.011 & 3.03 $\pm$ 0.45 & 3.55 $\pm$ 0.47 & 3.10 $\pm$ 0.53 & 2.22 $\pm$ 0.43 & --6.77 $\pm$ 0.85 & --1.63 $\pm$ 0.99\\
K029 & 0.301 $\pm$ 0.009 & 5.56 $\pm$ 0.34 & 3.00 $\pm$ 0.38 & 4.01 $\pm$ 0.42 & 1.26 $\pm$ 0.36 & --7.37 $\pm$ 0.80 & --0.48 $\pm$ 0.92\\
K033 & 0.129 $\pm$ 0.016 & 0.15 $\pm$ 0.71 & 2.31 $\pm$ 0.71 & 1.53 $\pm$ 0.85 & 3.78 $\pm$ 0.49 & 0.09 $\pm$ 0.91 & 2.21 $\pm$ 1.00\\
K036 & 0.309 $\pm$ 0.011 & 5.10 $\pm$ 0.44 & 4.77 $\pm$ 0.47 & 2.39 $\pm$ 0.56 & 2.62 $\pm$ 0.44 & --3.89 $\pm$ 0.90 & --9.75 $\pm$ 1.25\\
K040 & 0.166 $\pm$ 0.011 & 1.88 $\pm$ 0.46 & 3.35 $\pm$ 0.49 & 4.25 $\pm$ 0.53 & 1.49 $\pm$ 0.43 & --0.54 $\pm$ 0.75 & --4.86 $\pm$ 0.99\\
K043 & 0.260 $\pm$ 0.021 & 10.74 $\pm$ 0.60 & 7.59 $\pm$ 0.97 & 6.80 $\pm$ 0.89 & 2.96 $\pm$ 0.91 & --5.14 $\pm$ 0.84 & --4.36 $\pm$ 0.82\\
K044 & 0.052 $\pm$ 0.009 & 1.01 $\pm$ 0.36 & 1.55 $\pm$ 0.40 & 1.20 $\pm$ 0.46 & 1.83 $\pm$ 0.34 & --1.13 $\pm$ 0.58 & 3.30 $\pm$ 0.60\\
K051 & 0.111 $\pm$ 0.014 & 4.18 $\pm$ 0.54 & 2.60 $\pm$ 0.62 & 3.73 $\pm$ 0.69 & 1.84 $\pm$ 0.61 & 1.53 $\pm$ 0.86 & --9.10 $\pm$ 1.22\\
K052 & 0.030 $\pm$ 0.008 & 2.01 $\pm$ 0.32 & --0.44 $\pm$ 0.39 & --1.09 $\pm$ 0.45 & 2.57 $\pm$ 0.31 & 0.81 $\pm$ 0.50 & 3.23 $\pm$ 0.54\\
K056 & 0.057 $\pm$ 0.012 & --1.92 $\pm$ 0.53 & --0.17 $\pm$ 0.56 & 2.35 $\pm$ 0.59 & 1.50 $\pm$ 0.46 & --6.70 $\pm$ 0.88 & 0.66 $\pm$ 0.86\\
K057 & 0.236 $\pm$ 0.012 & 5.01 $\pm$ 0.46 & 2.24 $\pm$ 0.56 & 1.56 $\pm$ 0.65 & 3.52 $\pm$ 0.44 & --2.62 $\pm$ 0.96 & 0.90 $\pm$ 1.00\\
K061 & 0.193 $\pm$ 0.014 & 5.11 $\pm$ 0.50 & 0.73 $\pm$ 0.67 & 4.79 $\pm$ 0.65 & 2.52 $\pm$ 0.53 & 5.39 $\pm$ 1.00 & 6.04 $\pm$ 0.89\\
K065 & 0.219 $\pm$ 0.008 & 4.06 $\pm$ 0.31 & 3.99 $\pm$ 0.34 & 3.74 $\pm$ 0.38 & 1.75 $\pm$ 0.30 & --3.84 $\pm$ 0.60 & --0.92 $\pm$ 0.63\\
K067 & 0.077 $\pm$ 0.018 & 1.90 $\pm$ 0.71 & 5.66 $\pm$ 0.72 & --1.29 $\pm$ 0.98 & 1.28 $\pm$ 0.71 & --4.41 $\pm$ 1.26 & --1.12 $\pm$ 1.21\\
K069 & 0.203 $\pm$ 0.016 & 3.99 $\pm$ 0.62 & 2.66 $\pm$ 0.70 & 5.60 $\pm$ 0.75 & 4.09 $\pm$ 0.56 & --5.26 $\pm$ 1.40 & --8.82 $\pm$ 1.22\\
K074 & 0.121 $\pm$ 0.009 & 1.97 $\pm$ 0.36 & 0.23 $\pm$ 0.42 & 1.50 $\pm$ 0.46 & 1.23 $\pm$ 0.34 & 0.04 $\pm$ 0.56 & 2.73 $\pm$ 0.63\\
K078 & 0.122 $\pm$ 0.018 & --1.30 $\pm$ 0.82 & 7.86 $\pm$ 0.69 & 4.14 $\pm$ 0.90 & 1.78 $\pm$ 0.63 & --3.23 $\pm$ 1.23 & 1.10 $\pm$ 1.21\\
K083 & 0.177 $\pm$ 0.011 & 3.95 $\pm$ 0.43 & 1.60 $\pm$ 0.50 & 1.65 $\pm$ 0.55 & 2.03 $\pm$ 0.39 & --2.96 $\pm$ 0.86 & --2.97 $\pm$ 1.07\\
K085 & 0.132 $\pm$ 0.010 & 1.87 $\pm$ 0.42 & 1.98 $\pm$ 0.46 & 1.91 $\pm$ 0.51 & 1.04 $\pm$ 0.42 & 0.18 $\pm$ 0.71 & --3.07 $\pm$ 0.92\\
K086 & 0.095 $\pm$ 0.009 & 2.25 $\pm$ 0.36 & 0.51 $\pm$ 0.41 & 1.44 $\pm$ 0.47 & 1.64 $\pm$ 0.36 & --2.13 $\pm$ 0.62 & --1.24 $\pm$ 0.73\\
K089 & 0.143 $\pm$ 0.008 & 2.68 $\pm$ 0.32 & 2.08 $\pm$ 0.36 & 1.91 $\pm$ 0.40 & 3.11 $\pm$ 0.31 & --2.93 $\pm$ 0.60 & 0.62 $\pm$ 0.66\\
K092 & 0.119 $\pm$ 0.015 & 0.90 $\pm$ 0.62 & 0.96 $\pm$ 0.66 & 0.18 $\pm$ 0.79 & 1.40 $\pm$ 0.52 & 4.71 $\pm$ 0.84 & 2.77 $\pm$ 0.69\\
K109 & 0.086 $\pm$ 0.012 & 0.54 $\pm$ 0.51 & 3.50 $\pm$ 0.54 & 6.40 $\pm$ 0.57 & 0.37 $\pm$ 0.45 & --2.39 $\pm$ 0.71 & --8.18 $\pm$ 1.07\\
K110 & 0.053 $\pm$ 0.010 & 0.54 $\pm$ 0.39 & 0.53 $\pm$ 0.43 & 0.11 $\pm$ 0.49 & 3.31 $\pm$ 0.37 & --1.36 $\pm$ 0.77 & 2.72 $\pm$ 0.84\\
K113 & 0.262 $\pm$ 0.015 & 2.44 $\pm$ 0.62 & 3.37 $\pm$ 0.64 & 3.32 $\pm$ 0.73 & 3.54 $\pm$ 0.53 & --7.36 $\pm$ 1.05 & 1.42 $\pm$ 1.34\\
Metal--poor & 0.065 $\pm$ 0.004 & 1.52 $\pm$ 0.17 & 0.82 $\pm$ 0.19 & 1.13 $\pm$ 0.20 & 2.14 $\pm$ 0.19 & --1.31 $\pm$ 0.30 & 2.19 $\pm$ 0.28\\
Metal--rich & 0.213 $\pm$ 0.006 & 4.19 $\pm$ 0.21 & 3.09 $\pm$ 0.22 & 2.77 $\pm$ 0.24 & 2.13 $\pm$ 0.22 & --4.09 $\pm$ 0.49 & --2.52 $\pm$ 0.50\\
\hline
\end{tabular}
\label{tab:Lick}
\end{center}
\end{table*}

For this analysis, we have used the SSP models of 
\citeANP{Maraston00} (2000). These are our models of choice
since they cover the expected age--metallicity parameter space of
the NGC~524 GCs, and have also been shown to adequately reproduce the 
ages and metallicities of both Galactic globular clusters
\cite{Maraston00}, and LMC star clusters over a wide
age range \cite{Beasley02a}.
The evolutionary synthesis models of \citeANP{Maraston00} (2000) deal
with post-main sequence stellar evolution using the 
Fuel Consumption Theorem \cite{Renzini86}.
The TP-AGB phases are calibrated upon observations of star clusters
in the Magellanic Clouds, and adopt 
classical, non-overshooting stellar tracks \cite{Cassisi98}.
These models span a metallicity
range of --2.25 $\leq$ [Fe/H] $\leq$ +0.67, 
and an age range from 30 Myr to 15 Gyr, although here we only
consider models $\geq$ 0.5 Gyr.

\begin{figure*}
\centering
\centerline{\psfig{file=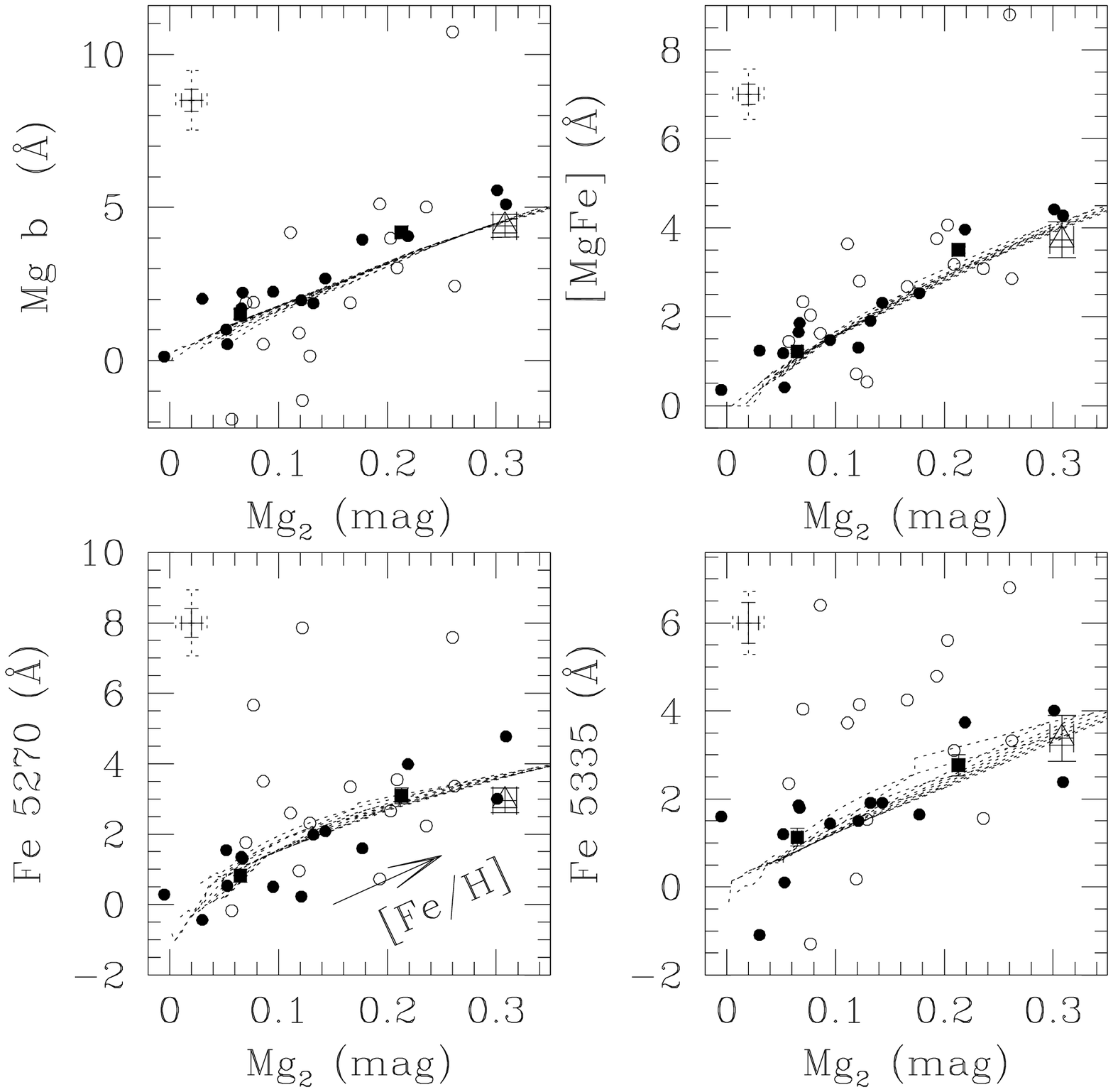,height=15cm}}
\caption{The NGC~524 GC data compared to the stellar population models of
Maraston \& Thomas (2000).
Small solid circles represent individual higher S/N GC data, the
open circles indicate lower S/N data. The average
uncertainties determined for the higher and lower S/N data are
shown by the solid and dashed error bars in the top-left corners
of the plots respectively.  
Solid squares with error bars indicate our composite data for the 
metal-rich ([Fe/H] $\geq$ --1.0) and metal-poor ([Fe/H] $<$
--1.0) clusters. The models are plotted such that they are
effectively degenerate in age, with metallicity increasing in the
direction of the arrow in the lower left panel. The open triangle
with error bars shows the nuclear line-strengths of NGC~524 taken
from Trager \etal (1998).}
\label{fig:ssp1}
\end{figure*} 

\begin{figure*}
\centering
\centerline{\psfig{file=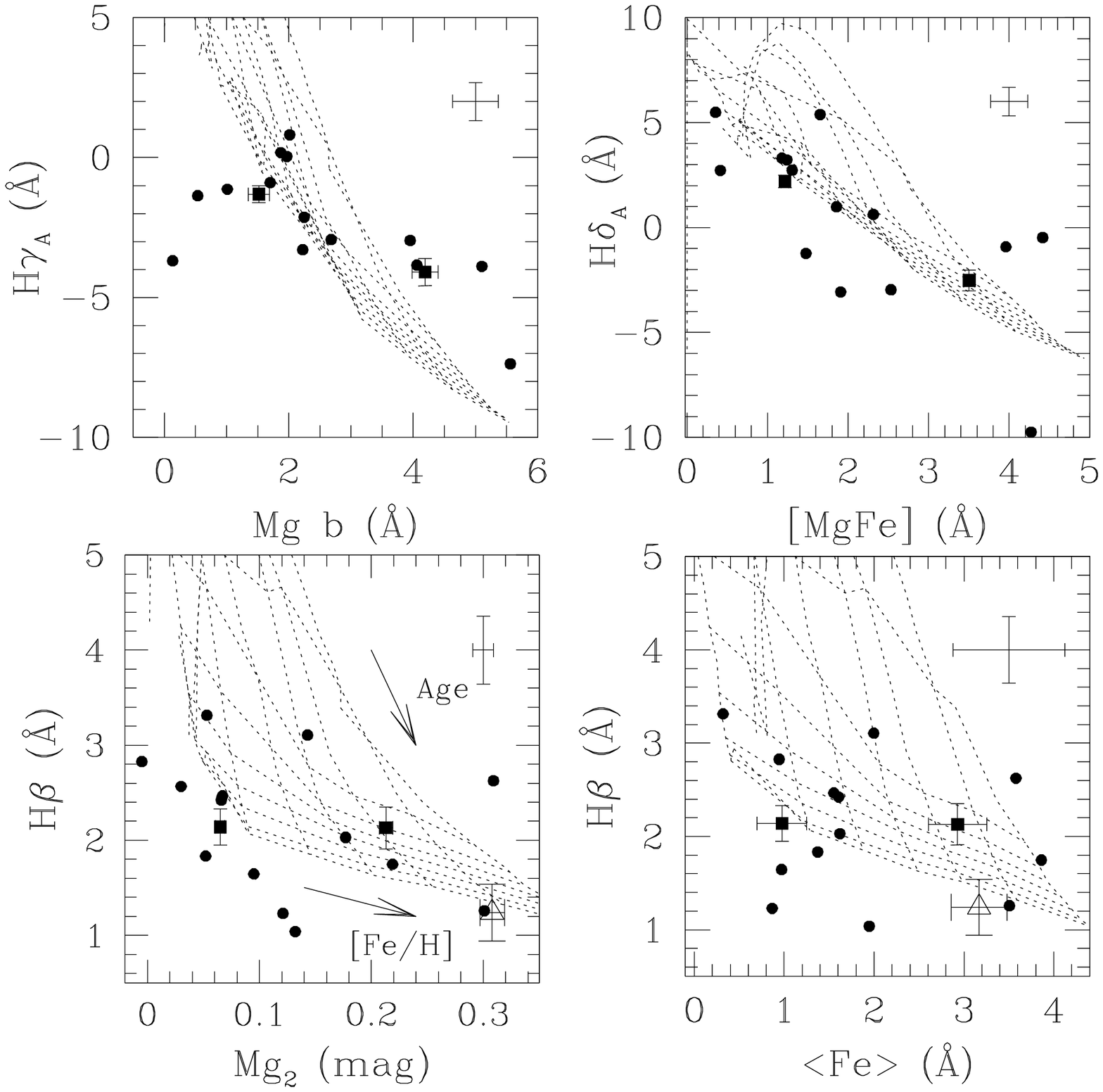,height=15cm}}
\caption{NGC~524 GC data compared to the stellar population models of
Maraston \& Thomas (2000). Grid ages are 1,3,5,7,9,11,13 \& 15
Gyr, metallicities are --2.25 --1.35, --0.84 (interpolated), 
--0.33, 0, 0.35, \& 0.67. 
Small solid circles represent individual higher S/N GC data, 
the average uncertainties determined for these data are 
shown by the error bars in the top-right corners of the plots. 
Solid squares indicate our binned data for the 
metal-rich ([Fe/H] $\geq$ --1.0) and metal-poor ([Fe/H] $<$
--1.0) clusters. The open triangle in the bottom two plots shows 
the nuclear line-strengths of NGC~524 from
Trager \etal (1998). The definitions of H$\gamma_{\rm A}$ and 
H$\delta_{\rm A}$ are given in Worthey \& Ottaviani (1997).}
\label{fig:ssp2}
\end{figure*} 

We have first compared different metallicity-sensitive Lick indices 
measured from our data with the \citeANP{Maraston00} SSP models.
This provides both a test of the SSP models' ability to describe the
data, and gives us an insight into the internal consistency of
these spectroscopic data.
This latter point is especially important since we have
no repeat observations of these clusters and, as discussed in
Section~\ref{Metallicities}, for relatively low
S/N spectroscopic data uncertainties in flat-fielding and 
sky-subtraction can become a dominant source of error.

We show in Figure~\ref{fig:ssp1} the NGC~524 data compared to the
SSP models for the commonly used Mg$_2$, Fe5270, Fe5335 and
the composite [MgFe] index defined by \citeANP{Gonzalez93} (1993).
To illustrate the importance of our error estimates when comparing to
the SSP models, we have split our spectroscopic sample into 
'high' S/N and 'low' S/N groups at S/N $\sim$ 15, 
yielding 14 GCs in the high S/N group. The cut at S/N $\sim$ 15 was
chosen such that the index error distributions were Gaussian
about their expected mean.
In Figure~\ref{fig:ssp1} these are represented by filled circles
(high S/N GCs) and open circles (low S/N GCs).

These data, shown in Figure~\ref{fig:ssp1}, exhibit varied 
behaviour depending upon the combination 
of indices used, but more importantly, upon the quality of the
data.
In all cases, not only does the higher S/N data show smaller
scatter (as expected, and indicated by the correspondingly smaller error
bars) but also exhibit Gaussian uncertainties, whereas the poorer
quality data shows indications of a bias towards larger index
values. This is particularly true of GC K043 which appears as 
an outlier in all four panels of Figure~\ref{fig:ssp1} and has 
the lowest S/N of the sample.

Concentrating solely on the higher S/N data, 
we find that the models (plotted such that they are effectively
degenerate in age and exhibit only a metallicity sequence) 
reproduce the data very well. This is also the case for 
our composite metal-poor and metal-rich GCs.
There is a suggestion of a systematic offset in the 
Mg {\it b} -- Mg$_2$ grid, since the composite GCs lie slightly 
above the grids, although the offset is not large.

For comparison purposes, 
we also show the position of the central nuclear line-strengths
of NGC~524 (through a standard 1.4$\arcsec \times $ 4$\arcsec$
aperture), taken from \citeANP{Trager98} (1998).
Two of the GCs in Figure~\ref{fig:ssp1} appear to have
metallicities similar to that of NGC~524 itself.
In principle, [$\alpha$/Fe] ratios may be derived from these
plots, but we defer further discussion of this subject to the next section.

For the rest of this analysis, we concentrate on the 14
GCs which have higher S/N. 
In Figure~\ref{fig:ssp2} we compare our data with the
\citeANP{Maraston00} (2000) models, for the $\langle$Fe$\rangle$
(Fe5270 + Fe5335)/2, Mg$_2$, H$\beta$, H$\gamma_A$ and 
H$\delta_A$ indices, and the combined [MgFe] index.
To assist in our interpretation of the grids, an additional 
metallicity interval at [Fe/H] = --0.84 has been included in
the models shown in Figure~\ref{fig:ssp2}, by linear interpolation
between the [Fe/H] = --1.35 and --0.33 lines.

It is clear that the individual data-points for the GCs populate
a large area of the SSP grid parameter space.
In the lower two panels of Figure~\ref{fig:ssp2} the distribution
in H$\beta$ is broadly consistent with the observational errors following
the old-aged loci of the grids.
The H$\delta_{\rm A}$ index also behaves similarly; however it is
clear that we can only measure H$\gamma_{\rm A}$ poorly.
We have also plotted the nuclear line-strengths of
NGC~524 itself, taken from \citeANP{Trager98} (1998), which 
indicate an old, metal-rich stellar system. 

To tighten constraints on the ages of the GCs, we show the
positions of our composite metal-poor and metal-rich GCs in 
Figure~\ref{fig:ssp2}. 
We find that from the H$\beta$ and H$\delta_A$ indices,
the metal rich composite GC appears somewhat younger than the 
metal-poor GC. The difference varies between the different age
diagnostics, but is in the range 2--5 Gyr at $\sim$ 2$\sigma$,
and is model dependent. Needless to say, this is consistent 
with the GCs being coeval (and old) in these current data.

In summary, the individual S/N of the GCs are generally too
low to place constraints upon individual GC ages.
Therefore we have co-added the individual spectra (separated
by metallicity) and find that these composite 
NGC~524 GCs have line-strengths consistent with old stellar
populations. The metal-poor and metal-rich sub-populations are
old and coeval at the 2$\sigma$ level of confidence.

\subsection{Abundance ratios}
\label{Abundances}

It has been known for some time that luminous elliptical 
galaxies exhibit non-solar abundance ratios from 
their integrated light (\eg \citeANP{OConnell76} 1976;
\citeANP{Peletier89} 1989; \citeANP{Worthey92} 1992). 
Integrated light studies also suggest that GCs associated with
both ellipticals (NGC~1399: \citeANP{KisslerPatig98} 1998; 
\citeANP{Forbes01} 2001) and spirals 
(M~31: \citeANP{deFreitasPacheco97} 1997; 
the Sombrero: \citeANP{Larsen02} 2002; the Milky Way: 
\citeANP{Borges95} 1995 and see later this section) 
also exhibit similar abundance patterns.\footnote{Note that Trager
\etal (2000a) and several other authors have argued that [Mg/Fe] $>$0 
should be viewed as a suppression in Fe rather than any real
``$\alpha$-element enhancement''}

\begin{figure}
\centering
\centerline{\psfig{file=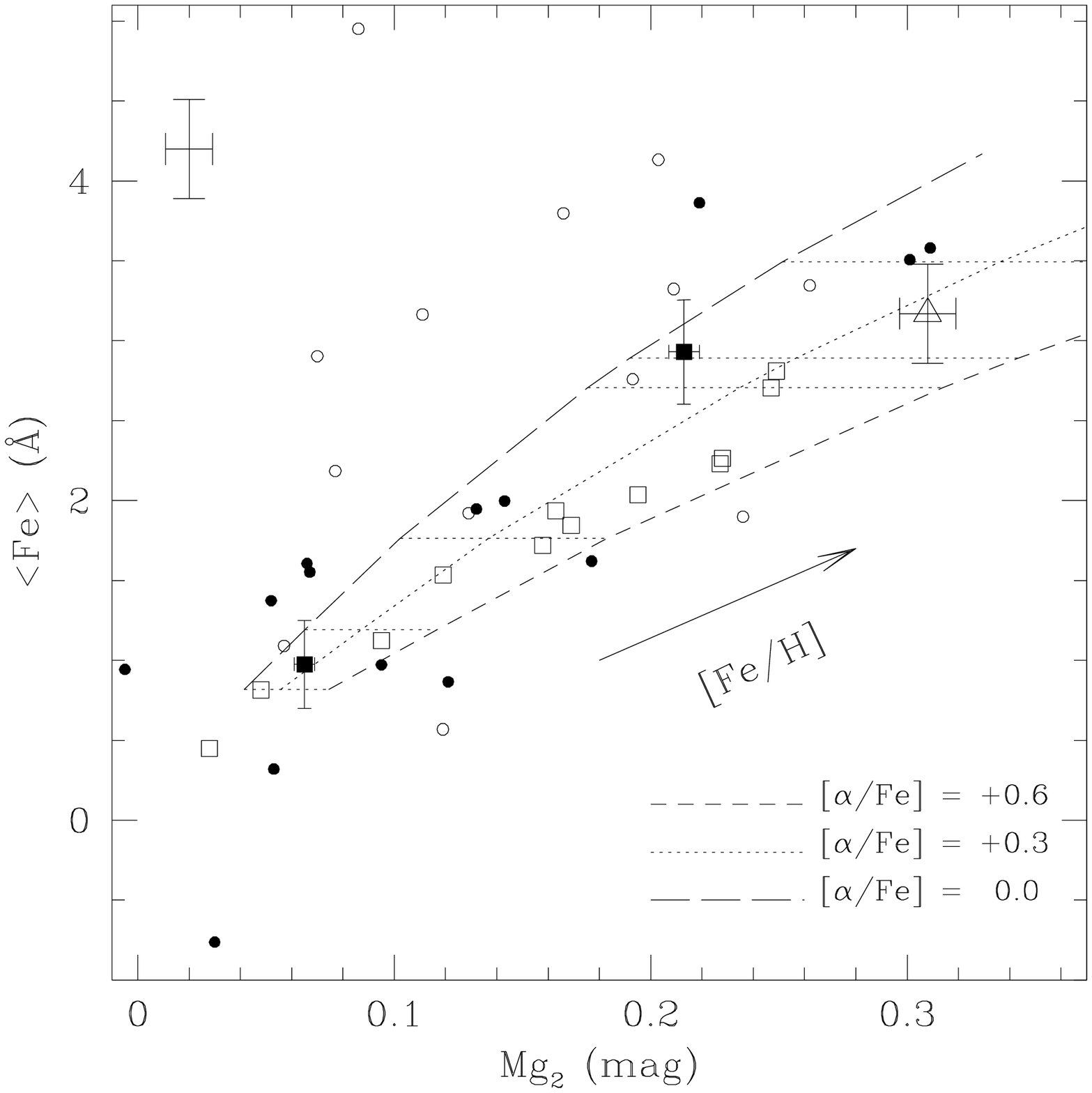,height=9cm}}
\caption{Comparison of NGC~524 GCs and Galactic GCs with 
the $\alpha$-enhanced models of
Milone \etal (2000). The horizontal dotted lines indicate 
iso-metallicity lines of [Fe/H] = --1.8, --1.3, --0.8,
--0.2, --0.1, \& +0.2 (left to right).
Our NGC~524 data are shown
as small solid circles (for the high S/N group). 
The average errors for these data are given
by the error bars in the upper-left corner.
The positions of our composite spectra with
their associated boot-strapped uncertainties are shown as solid squares.
The Galactic GCs (data from Cohen \etal 1998) are indicated by
open squares, and possess mean errors of 0.01 mag in Mg$_2$ and
0.15 \AA\ in $\langle$Fe$\rangle$.
The open triangle indicates the position of the
nucleus of NGC~524 taken from Trager (1998).
Unlike the Galactic GCs, the NGC~524 data exhibit a wide range of 
[$\alpha$/Fe] ratios, with a weak trend of decreasing
[$\alpha$/Fe] with increasing [Fe/H].}
\label{fig:milone}
\end{figure}

To assess the [$\alpha$/Fe] ratios in our data, following
\citeANP{Larsen02} (2002) we have used the ``$\alpha$-enhanced''
models of \citeANP{Milone00} (2000), who present a calibration of
the Mg$_2$ and TiO ($\lambda\lambda$7464) indices 
(the latter not measured in our
spectra), as a function of [Fe/H] and [$\alpha$/Fe] for SSP
models. These models rely on the fact that the Mg$_2$ and 
$\langle$Fe$\rangle$ indices should trace [Mg/H] and 
[Fe/H] differently, thereby providing a diagnostic of
any departures in [$\alpha$/Fe] from solar values.

Prior to deriving [$\alpha$/Fe] ratios for the NGC~524
GCs, we have performed a consistency check of the 
\citeANP{Milone00} (2000) calibrations by  comparing 
them to stellar populations with independently derived 
[$\alpha$/Fe] ratios, namely Galactic GCs. 
This is a necessary verification since the test applied 
by \citeANP{Milone00} (2000) themselves (using the
$\langle$Fe$\rangle$, Mg$_2$, and TiO indices of 12 elliptical galaxies)  
only covered the narrow metallicity range of --0.2 $\leq$ [Fe/H]
$\leq$ +0.2.\footnote{We note also that \citeANP{Borges95} (1995)
performed a similar comparison between their ``$\alpha$-enhanced''
SSP models and the Galactic GC data of \citeANP{Burstein84}
(1984), although over smaller metallicity range and with poorer quality data.}

To this end, we use the Galactic GC data from \citeANP{Cohen98}
(1998), who have scanned the cores of 12 Galactic GCs in the 
metallicity range --2.3 $\leq$ [Fe/H] $\leq$ 0, achieving 
a typical S/N of $\sim$ 500. 
\citeANP{Cohen98} (1998) took these data with LRIS on Keck I, 
using an essentially identical set-up to the NGC~524 observations
presented here. 
This should ensure that any instrument-dependent
differences are minimised in our comparisons.
However, \citeANP{Cohen98} (1998) only measured a subset of 
Lick indices for these data, and did not include Fe5335 or
Mg$_2$ in their analysis. For the purposes of our analysis, 
we have re-measured these Lick indices (along with their
associated uncertainty) from the LRIS spectra
(kindly supplied by J. Blakeslee), broadened to the Lick/IDS
resolution. From a comparison with the six
clusters in common with \citeANP{Burstein84} (1984), we find no
evidence for a significant offset in the Mg$_2$ index between these data and
the Lick system, contrary to the claim by \citeANP{Cohen98} (1998).

We compare the Galactic GC data to the \citeANP{Milone00} (2000)
models in Figure~\ref{fig:milone}. The majority of the Galactic GCs
in the figure lie in the 0.3 $\leq$ [$\alpha$/Fe] $\leq$ 0.6 
range of the models, suggesting that these Galactic GCs exhibit 
non-solar abundance ratios from their integrated spectra. 
Indeed, the Galactic GCs seemingly trace a locus which extends
to the ``$\alpha$-enhanced'' metal-rich position of the nucleus of
NGC~524 itself.

We list the derived [$\alpha$/Fe] ratios for these Galactic GCs
using the \citeANP{Milone00} (2000) models in column 4 of 
Table~\ref{tab:abundances}. For comparison, 
we also list literature [Fe/H] values of the Galactic
GCs, in addition to the extant [$\alpha$/Fe] ratios of GCs 
obtained from high-resolution spectroscopy
of individual cluster giants (from the compilation of 
\citeANP{Carney96} 1996; supplemented with data from
\citeANP{King98} 1998; \citeANP{Cohen99} 1999
and \citeANP{Carretta01} 2001).
These literature [$\alpha$/Fe] ratios derive from
the unweighted mean of three $\alpha$-elements : 
[$\langle$Si + Ca + Ti$\rangle$/Fe]. As suggested by 
\citeANP{Carney96} (1996), these elements should give a
good indication of ``primordial'' $\alpha$-element abundances,
since these elements are neither synthesised nor destroyed on the
red giant branch in globular clusters.

We find that the agreement between [$\alpha$/Fe] ratios
derived from the integrated spectra and 
high-resolution measurements is generally good.
We note that from their integrated spectra,  
the abundance ratios of the Galactic GCs remain 
roughly constant at [$\alpha$/Fe] $\sim$ + 0.4, 
from [Fe/H] = --1.8, up to solar metallicities.
The Galactic GC M13 appears to have an [$\alpha$/Fe] ratio 
somewhat closer to the solar abundance. 
M92 appears too metal-poor for the \citeANP{Milone00} (2000)
models and we do not discuss it further.
In general, the predictions of the \citeANP{Milone00} (2000)
models for the Galactic GC [Fe/H] values are consistent with the
literature values (Table~\ref{tab:abundances}). 
However, the models overestimate the metallicity of 
M13 by $\sim$ 0.3 dex. This is potentially the origin of
the lower [$\alpha$/Fe] ratios obtained for this GC.
We also note that at these metallicities
the \citeANP{Milone00} (2000) models have a small range and 
abundance ratio discrimination is lost.\footnote{An alternative 
explanation is that we are
seeing the effects of altered surface abundances of magnesium due
to deep mixing in metal-poor globular cluster giants.
\citeANP{Kraft93} (1993) find a strong O/Na anti-correlation in
nine M13 giant stars.}

We conclude that the \citeANP{Milone00} (2000) 
models are able to reproduce the abundance ratios of 
Galactic GCs from integrated spectra over our observed
metallicity range reasonably well\nocite{Zinn85}

\begin{table}
\begin{center}
\caption{[$\alpha$/Fe] ratios of Galactic Globular
Clusters. Column 2 gives the cluster metallicity derived from 
the Milone \etal (2000) SSP models, Column 3 gives the literature 
cluster [Fe/H] on the Zinn \& West (1985) scale (Harris 1996), 
Column 4 lists our derived abundance ratios from the SSP models, 
Column 5 lists the available literature values and the sources of
the values in Column 5 are given in Column 6.}
\begin{tabular}{lcccccc}
\hline
\hline
ID & [Fe/H] & [Fe/H]& [$\alpha$/Fe] & $\langle[\alpha$/Fe]$\rangle$ & Ref\\
	& (SSP)&(Lit)&(SSP)&(Lit)&\\
\hline
M92 	  & ---    &  -2.29 & ---            &  +0.24$\pm$0.13$^a$ & 1\\
M13  	  & -1.80 &  -1.54 & +0.15$\pm$0.20 &  +0.29$\pm$0.02 & 2\\
M4   	  & -1.41 &  -1.20 & +0.45$\pm$0.14 &  +0.30$\pm$0.03 & 3,4\\	
N6171  & -1.00 &  -1.04 & +0.32$\pm$0.12 &  ---		   &\\
M71	  & -0.85 &  -0.73 & +0.46$\pm$0.12 &  +0.31$\pm$0.02 & 3,5\\   
N6539  & -0.65 &  -0.66 & +0.45$\pm$0.12 &  ---		   &\\
N6760  & -0.44  &  -0.52 & +0.50$\pm$0.10 &  ---		   &\\
N6356  & -0.75 &  -0.50 & +0.45$\pm$0.10 &  ---		   &\\
N6624  & -0.70 &  -0.42 & +0.40$\pm$0.07 &  ---		   &\\
N6440  & -0.47 &  -0.34 & +0.50$\pm$0.07 &  ---		   &\\
N6553  & -0.20 &  -0.25 & +0.30$\pm$0.05 &  +0.20$\pm$0.06 & 6\\
N6528  & -0.14 &  -0.17 & +0.40$\pm$0.04 &  +0.30$\pm$0.09$^b$ & 7\\
\hline
\end{tabular}
\noindent Sources: 1 : \citeANP{King98} (1998); 2 :
\citeANP{Kraft93} (1993); 3 : \citeANP{Gratton86} (1986); 4:
\citeANP{Brown92} (1992); 5 : \citeANP{Sneden94} (1994);  
6 : \citeANP{Cohen99} (1999); 
7 : \citeANP{Carretta01} (2001). \noindent$^a$ : shows
star-to-star abundance variations, [$\alpha$/Fe] derived from 
Ti + Ca only. $^b$ : [$\alpha$/Fe] derived from Si + Ca only.
\label{tab:abundances}
\end{center}
\end{table} 

In Figure~\ref{fig:milone} we show the position of the individual
NGC~524 GCs compared to the \citeANP{Milone00} (2000) models.
The uncertainties on these points are relatively large, as
indicated by the mean error bar shown in the plot. 
They do, however, show some
trend in that their [$\alpha$/Fe] ratios appear to weakly 
decrease as metallicity increases.
We also show the location of our two composite NGC~524 GCs in the
figure compared to the models. 
These composite data behave similarly to the individual
datapoints; the metal-rich bin shows a lower [$\alpha$/Fe] ratio
than the metal-poor bin.
Similar to the situation for the Galactic GCs (and for 
the nucleus of NGC~524 itself), the 
location of these data suggests that many of the NGC~524 GCs
are $\alpha$-enhanced. The metal-poor bin possesses 
[$\alpha$/Fe] = 0.3$\pm$0.3, whilst the metal-rich
bin exhibits [$\alpha$/Fe] = 0.10$\pm$0.15.

The non-solar [$\alpha$/Fe] ratios seen in luminous ellipticals
are generally thought to be a result of an ISM
largely polluted by type-II supernovae,
for which viable explanations include 
either relatively short star-formation time-scales  and/or
an IMF skewed towards more massive stars 
(\eg  \citeANP{Wyse88} 1988; \citeANP{Yoshii96} 1996).
Abundance ratio differences between the NGC~524 
GC sub-populations suggest that they either formed
in sites with different enrichments, 
or over different timescales.
Interestingly, our data hints that the metal-rich clusters may 
be more iron-enriched (i.e. from type-I products) 
than the bulk of the spheroid of NGC~524. 

It is currently not clear whether galaxy nuclei and GCs 
undergo similar chemical enrichment, and 
indeed the interpretation of such abundance variations are 
not unambiguous from low-dispersion analysis.
Existing data suggests that the abundance ratios of 
alpha to iron-peak elements show complex behaviour 
which may differ between GCs and galaxies. 
For example, [Ca/Fe] $>$ 0 in Milky Way GCs 
(\eg \citeANP{Carney96} 1996), whilst 
luminous ellipticals apparently have 
[Ca/Fe] $\simeq$ 0 (\eg \citeANP{Worthey98} 1998), 
or may in fact be underabundant (\citeANP{Saglia02} 2002; 
\citeANP{Cenarro03} 2003). 
Whether calcium in elliptical galaxies is truly suppressed 
with respect to other $\alpha$-elements, or that this 
apparent suppression is due to deficiencies in current stellar 
population models (\eg see discussion in \citeANP{Borges95} 1995),
requires further investigation. 

\section{Radial Velocities of the Globular Clusters}
\label{Radial}

\subsection{Kinematics of the Cluster System}
\label{Kinematics}

\begin{figure}
\centering
\centerline{\psfig{file=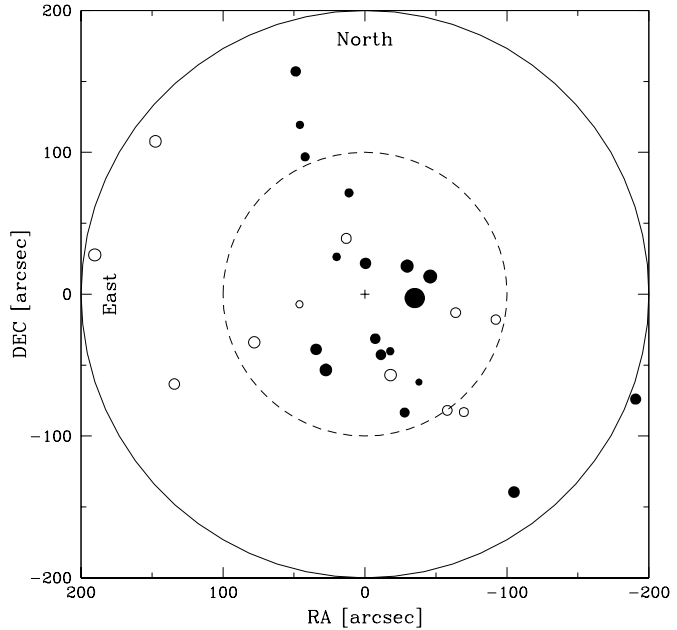,height=9cm}}
\caption{The spatial distribution of our spectroscopic GC
sample.  Open symbols represent approaching GCs,
solid symbols represent receding GCs with respect to our derived
systemic velocity ($2432\pm38 \kms$). Symbol size corresponds to
the GC velocity difference from this systemic
velocity. Concentric circles indicate 1 and 2 effective radii 
respectively ($\sim$ 13 kpc and 26 kpc respectively). }
\label{fig:velocities}
\end{figure}

We have also examined the kinematics of these NGC~524 data
using our measured radial velocities. Whilst our sample is not
large enough to perform a comprehensive kinematical analysis
(e.g. \citeANP{Zepf00} 2000; \citeANP{Cote01}
2001; \citeANP{Cote03} 2003), we have been able to place the 
first constraints on the dynamics of the NGC~524 GC system.

In Figure~\ref{fig:velocities}, we show the spatial and velocity
distribution of our whole spectroscopic sample with respect to 
NGC~524. The dashed and solid concentric circles indicate 
approximately 1 and 2 effective radii of the galaxy respectively
($\sim$ 13 kpc and 26 kpc respectively).
From our entire cluster sample (29 clusters) we derive a mean 
velocity of $2455\pm 44\kms$, in good agreement with the literature
systemic velocity of NGC~524 of $2421\pm19 \kms$
\cite{RC3}. 
Removing an extreme cluster (\#K040, $v=3141\pm71 \kms$) 
from the sample, we obtain a mean velocity
of $2432\pm38 \kms$, in even better agreement with the literature.

To obtain the velocity dispersion of the full sample, we have
used a maximum likelihood estimator \cite{Pryor93}.
We obtain a velocity dispersion of $225\pm33 \kms$ for the 
entire cluster sample, at a mean radius of 89 arcsec 
($\sim$ 12 kpc) from the galaxy centre. Removing the outlying cluster,
we obtain a velocity dispersion of $186\pm29 \kms$.
We note that a measurement of the stellar velocity dispersion 
at 40 arcsec is reported by \citeANP{Simien00} (2000) to be 
around this value (between 150 and 210 $\kms$ depending on
the side of the galaxy).

We have looked for rotation in the entire cluster system by
fitting a sinusoid of the general form:

\begin{equation}
V(r)=V_{\rm rot} sin(\theta-\theta_0) + V_0
\label{eq:rotation}
\end{equation}

\noindent to the position angle and GC velocity data using a
nonlinear, least-squares fit (\eg \citeANP{Sharples98} 1998). 
$\theta_0$ and V$_{\rm rot}$ are free parameters, and represent 
the position angle of the line of nodes and the rotation velocity
of the GCs respectively. V$_0$ is the systemic velocity of the GC
system (\ie our derived mean velocity).

\begin{figure}
\centering
\centerline{\psfig{file=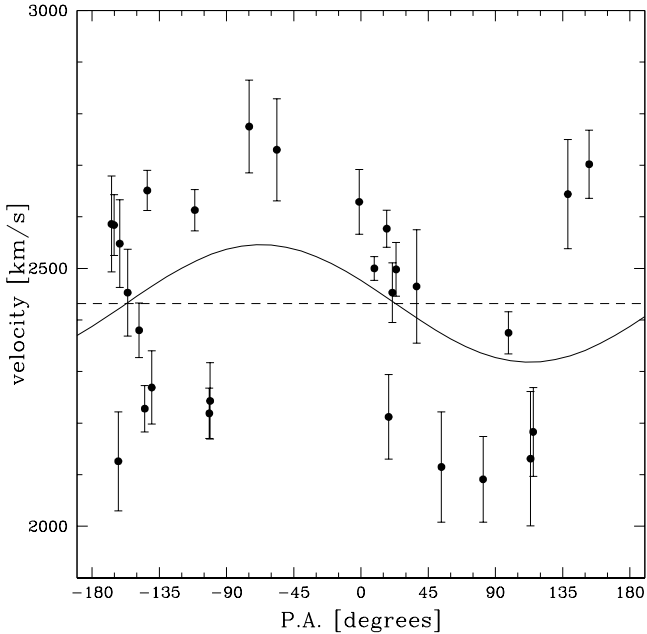,height=9cm}}
\caption{Plot of the velocities of NGC~524 globular clusters 
against their position angle (measured north through east). 
Our best-fit sinusoid is shown, yielding rotation of $114\pm60 \kms$
around a position angle of $22\pm 27 \deg$. There is a clear
signature of rotation, even with our relatively small number GCs.}
\label{fig:rotation}
\end{figure}

We find that the GC system shows rotation of $114\pm60 \kms$
(excluding the outlying cluster), around a position angle of 
$22\pm 27 \deg$. We plot the radial velocities of these clusters
against position angle, along with our best (weighted) sinusoidal fit in 
Figure~\ref{fig:rotation}.
Interestingly, the rotation we derive is similar to the maximum observed 
rotation velocity of the stellar light of $124\pm7 \kms$.
However this measurement is {\it along} a position angle of $38
\deg$ \cite{Simien00}. Therefore, the stellar rotation
appears to be at $\sim 180 \deg$ to the GCs if the position angle
conventions are correct.

Many luminous early-type galaxies host at least two
populations of GCs, as suggested by their bimodal colour
distributions (\eg \citeANP{Gebhardt99} 1999; \citeANP{Kundu01}
2001; \citeANP{Larsen01} 2001) and increasingly by their
differing kinematic properties (\eg \citeANP{Cohen97} 1997;
\citeANP{Zepf00} 2000; \citeANP{Cote01} 2001; \citeANP{Geisler01}
2001).
NGC~524 also possesses a GC system with a bimodal colour
distribution (\eg \citeANP{Larsen01} 2001 and see 
Figure~\ref{fig:colours}). We have looked at the kinematics of 
these sub-populations in our present sample.

We have separated the cluster sample into metal-poor ([Fe/H] $<$
--1.0) and metal-rich ([Fe/H] $\geq$ --1) groups
on the basis of the metallicities derived in Section~\ref{Metallicities}.
Performing the same analysis as described above, we find that the
metal-poor GCs (17 clusters) dominate the rotation in our 
sample, with a rotation of $147\pm 75 \kms$ around a position angle of
$6\pm 26 \deg$. In contrast, we find that the metal-rich GCs 
(11 clusters excluding the outlier) show no significant
rotation, with $68\pm84 \kms$.
Although the rotation of the two metallicity groups are not
strongly constrained at this point, this result seems to differ
from that of the NGC~3115, the nearby lenticular galaxy studied
by \citeANP{Kuntschner02} (2002). From a total sample of 24 GCs,
\citeANP{Kuntschner02} (2002) found that {\it both} the
metal-poor and metal-rich GCs had a similar rotation signal. 

Returning to NGC~524, the velocity dispersions of the two cluster populations, 
when neglecting rotation, differs only slightly: $197\pm39 \kms$ 
for the metal-poor GCs, $169\pm43 \kms$ for the metal-rich 
GC sub-population. 
Moreover, when the rotational component is taken into account, the
dispersion of the two sub-populations are very similar (201
$\kms$ versus 193 $\kms$). 
The kinematical results for the NGC~524 GC system 
are summarised in Table~\ref{tab:kinematics}.

The lack of rotation in the metal-rich population translates into
a rather low value for V$_{\rm rot}$/$\sigma$, an often used
diagnostic for the degree of rotation over anisotropy.
For the metal-rich clusters we find 
V$_{\rm rot}$/$\sigma$ = 0.40 $\pm$ 0.64,
where the rather large error reflects the small sample of
clusters. This is similar to the situation found by
\citeANP{Zepf00} (2000) for the NGC~4472 GC system, although they
also found only a modest level of rotation in the metal-poor clusters.

\begin{table*}
\begin{center}
\caption{The kinematic properties of the NGC~524 globular cluster
system.}
\begin{tabular}{lccccccc}
\hline
\hline
Sample 	& N & V$_0$ & V$_{\rm rot}$ &$\theta_0$ & $\sigma$ &V$_{\rm rot}$/$\sigma$\\
	&   & ($\kms$) & ($\kms$) &(degrees) & ($\kms$) & &\\ 
\hline
All clusters excluding \#K040	&28 & 2432$\pm$38 & 114$\pm$60 & 22$\pm$27 & 186$\pm$29 & 0.61$\pm$0.17\\
All Metal-poor clusters		&17 & 2415$\pm$54 & 147$\pm$75 & 6$\pm$26  & 197$\pm$39 & 0.75$\pm$0.22\\
Metal-rich clusters excluding \#K040 &11 & 2447$\pm$60 &  68$\pm$84 & ---	   & 169$\pm$43 & 0.40$\pm$0.64\\
\hline
\end{tabular}
\label{tab:kinematics}
\end{center}
\end{table*} 

These results hint at either different density profiles or 
different families of orbits for the GC sub-populations
(\eg see \citeANP{KisslerPatig98a} 1998 for a similar situation in M87). 
The most likely explanation is that this is due to the 
metal-poor clusters having a flatter density distribution than
the metal-rich clusters, as observed in luminous galaxies
(\eg see \citeANP{Rhode01} 2001 for the specific case of NGC~4472).

However, it does remain puzzling why the metal-poor globular clusters
rotate at a right angle with respect to the stellar light of NGC~524. 
Integral field data available for this galaxy 
(see \citeANP{deZeeuw02} 2002) will clarify
whether or not it is connected to the kinematically decoupled core.
Clearly a larger sample of radial velocities for the NGC~524 GCs
is desirable for a more detailed analysis of these issues.

\subsection{The Mass of NGC~524}
\label{Mass}

Globular clusters can be used to good effect
as test-particles to probe the 
gravitational potential of their parent galaxies (\eg
\citeANP{Cohen97} 1997; \citeANP{KisslerPatig98} 1998; 
\citeANP{Zepf00} 2000), and are 
complementary to other techniques such as studies of planetary
nebulae (\eg \citeANP{Hui95} 1995; \citeANP{Arnaboldi98} 1998)
and integrated light (\eg \citeANP{Dressler84} 1984;
\citeANP{Bender94} 1994).

\begin{figure}
\centering
\centerline{\psfig{file=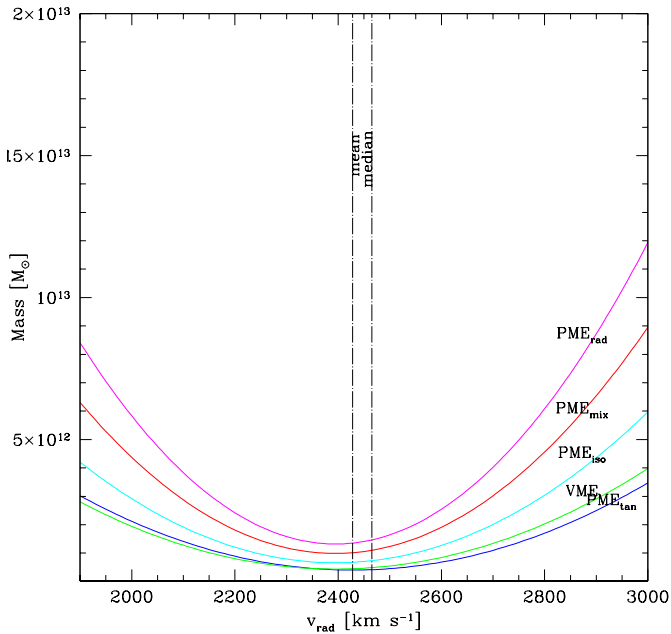,height=9cm}}
\caption{Solutions for the mass of NGC~524 from our sample of
globular clusters. VME = Virial mass estimator, PME = projected mass 
estimator. For the PME, we make several assumptions about the GC
orbits, radial (rad), a mixture of radial and isotropic (mix), 
isotropic (iso) and tangential (tan).}
\label{fig:mass}
\end{figure}

Our present spectroscopic sample allows us to 
estimate the total mass of the galaxy within a radius
of $\sim$ 200\arcsec, \ie within roughly 2 effective radii
($\sim$ 26 kpc) of NGC~524.
For this purpose we used the virial (VME) and projected 
(PME) mass estimators, as described in \citeANP{Bahcall81} 
(1981) and \citeANP{Heisler85} (1985) respectively.
With our adopted distance to NGC 524 of 28.2 Mpc \cite{RC3}, 
we obtain masses of $\sim 4\times 10^{11}$ M$_\odot$ (VME) and 
between $4\times 10^{11}$ (PME, purely tangential orbits) and 
$13\times 10^{11}$ M$_\odot$ (PME, purely radial orbits).
We show in Figure~\ref{fig:mass} the various mass-solutions we
obtain for this galaxy, resulting from our assumptions about the
nature of the NGC~524 cluster orbits.

\section{Summary and Conclusions}
\label{Conclusions} 

We have obtained low-resolution spectra for 41 GC candidates 
associated with the lenticular galaxy NGC~524. From this
sample, 29 candidates are identified as genuine GCs on the basis of
their radial velocities. Deriving mean metallicities
for the NGC~524 GCs, we find our sample spans a wide range in
metallicity, with --2.0 $\leq$ [Fe/H] $\leq$ 0.

The individual S/N of our spectra are generally insufficient 
to derive useful age constraints, since we rely on age discrimination
to come from individual indices (\eg H$\beta$). Therefore  
we have co-added the GCs into metal-poor ([Fe/H] $<$ --1.0)
and metal-rich ([Fe/H] $\geq$ --1.0) 'composite' GCs.
From comparison with the stellar population models 
of \citeANP{Maraston00} (2000), we find that the composite metal-rich
and metal-poor GC sub-populations both appear old, and are coeval
within the 2 $\sigma$ uncertainties.

We have examined the abundance ratios of the NGC~524 GCs
using the $\alpha$-enhanced stellar population models of
\citeANP{Milone00} (2000).
The calibration of these models was first tested using a sample of 
high S/N Galactic GC integrated spectra \cite{Cohen98}, 
which have independently determined [$\alpha$/Fe] ratios
from high resolution spectroscopy.
We find that the model predictions are in good agreement with these
literature values.
Comparing the \citeANP{Milone00} (2000) models to our 
data,  we find a weak trend of decreasing [$\alpha$/Fe]
with increasing [Fe/H].
This is supported by the co-added data, with the composite
metal-poor GC possessing [$\alpha$/Fe] $\sim$ 0.3, whereas the
metal-rich composite GC shows [$\alpha$/Fe] $\sim$ 0.1. 
The lower [$\alpha$/Fe] ratios of the metal-rich clusters 
may in fact reflect a mix of [$\alpha$/Fe] ratios 
(\ie sub-populations amongst the metal-rich clusters)
as found by \citeANP{Kuntschner02} (2002) for the lenticular
galaxy NGC~3115. 

We have also investigated the kinematics of the NGC~524 GC
system.
After the removal of one outlying GC, we obtain a velocity dispersion 
of $186\pm29 \kms$.
The entire cluster system shows a rotation of $114\pm60 \kms$
(excluding the outlying cluster), around a position angle of 
$22\pm27 \deg$.
By separating the GC systems into metal-rich and metal-poor
components (at [Fe/H]=--1.0), we find that the NGC~524 GC
sub-populations potentially exhibit different kinematics.  
Both sub-populations have similar 
(neglecting rotation) velocity dispersions 
($197\pm40 \kms$ and $169\pm 49 \kms$
respectively), but the metal-poor clusters show signs of rotation
($147\pm75 \kms$), whereas the metal-rich clusters do not
($68\pm84 \kms$).

Finally, using the entire GC system, we derive a virial and projected 
mass estimation for NGC~524 of between 4 $\times$ 10$^{11}$ $\Msun$
and 13 $\times$ 10$^{11}$ $\Msun$ (depending on the assumed 
orbital distribution) interior to $\sim$ 2 effective radii of this galaxy.

\section{Acknowledgements}

We thank Soeren Larsen and Michael Pierce for useful comments and
suggestions, John Blakeslee for supplying the Galactic GC data
and the anonymous referee, who greatly improved the presentation of the
paper, and noticed an error in the original manuscript.
Part of this research was funded by NSF grant AST 9900732 and
AST-0206139.
The data presented herein were obtained at the
W.M. Keck Observatory, which is operated as a scientific partnership among
the California Institute of Technology, the University of California and
the National Aeronautics and Space Administration.  The Observatory was
made possible by the generous financial support of the W.M. Keck
Foundation. This research has made use of the NASA/IPAC Extragalactic
Database (NED), which is operated by the Jet Propulsion Laboratory,
Caltech, under contract with the National Aeronautics and Space
Administration. 


\end{document}